\documentclass[12pt]{article}

\usepackage{pdfpages}

\usepackage{etoolbox}
\AtBeginEnvironment{quote}{\par\singlespacing\ttfamily\hyphenchar\font=`\-\spaceskip=.4em plus .4em\xspaceskip=.4em}
\usepackage{array,tikz}
\usetikzlibrary{tikzmark,arrows.meta}

\usepackage{amsfonts}

\usepackage[letterpaper, portrait, margin=1in]{geometry}
\usepackage{dsfont}
\usepackage{nicefrac,xfrac}
\usepackage{setspace}
\usepackage{indentfirst}
\usepackage{natbib}
\usepackage{mathtools}
\usepackage{adjustbox}
\usepackage[capposition=top]{floatrow}
\DeclarePairedDelimiter\ceil{\lceil}{\rceil}
\DeclarePairedDelimiter\floor{\lfloor}{\rfloor}

\PassOptionsToPackage{dvipsnames,table}{xcolor}
\usepackage[dvipsnames]{xcolor}
\usepackage{booktabs}
\usepackage{pifont}
\usetikzlibrary{positioning, fit}
\newcommand{\ra}[1]{\renewcommand{\arraystretch}{#1}}

\usepackage[shortlabels]{enumitem}
\usepackage{multirow}
\usepackage[multiple]{footmisc}
\usepackage{hyperref,graphicx,import,amsmath}
\hypersetup{colorlinks,citecolor=blue,linkcolor=blue,urlcolor=blue}
\providecommand{\tabularnewline}{\\}
\definecolor{myred}{RGB}{141,22,24}
\usepackage{verbatim}
\usepackage{bm}
\usepackage{upgreek}

\def\signed #1{{\leavevmode\unskip\nobreak\hfil\penalty50\hskip2em
  \hbox{}\nobreak\hfil(#1)%
  \parfillskip=0pt \finalhyphendemerits=0 \endgraf}}

\newsavebox\mybox

\usepackage{subcaption}

\newenvironment{head}
  {\par\setlength{\abovedisplayskip}{1pt}\setlength{\belowdisplayskip}{1pt}\setlength{\leftskip}{1cm}\setlength{\rightskip}{1cm}\noindent\ignorespaces}
  {\par}

\graphicspath{ {images/} }
\usepackage[font=footnotesize]{caption}

\usepackage{amsthm}

\newtheorem{theorem}{Theorem}
\newtheorem{corollary}{Corollary}

\newtheorem{example}{Example}
\newtheorem{lemma}{Lemma}
\newtheorem{remark}{Remark}

\newcommand{\bias}{\mathsf{bias}}
\newcommand{\uni}{\mathrm{u}}
\newcommand{\tot}{\mathrm{tot}}

\title{Affirmative Action's Cumulative Fractional Assignments\thanks{We are deeply grateful to Tayfun S\"{o}nmez, M. Utku \"{U}nver, and M. Bumin Yenmez for their invaluable guidance and continuous support. We wish to thank the Editor (Scott Kominers), three anonymous reviewers, Abhijit Banerji, In\'{a}cio B\'{o}, Umut Dur, Mehmet Ekmekci, Xiang Han, Gaoji Hu, Hideo Konishi, Morimitsu Kurino, Debasis Mishra, Thayer Morrill, William Phan, Uzi Segal, Rohit Vaish, Rakesh Vohra, Yongchao Zhang, and seminar participants at North Carolina State University, Boston College, the 2022 SAET Conference at the Australian National University, Shenzhen University Public Policy and Experimental Economics Workshop, Indian Institute of Technology - Delhi, Indian Statistical Institute - Delhi, Shanghai University of Finance and Economics, the 2024 AMES in Hangzhou, China, the 2024 Conference on Mechanism and Institution Design at Corvinus University of Budapest, and Delhi School of Economics for helpful comments and discussions. We are grateful to Yuan Gao, Xi Jin, and Bozhang Xia for helpful research assistance. This project began in 2019 after Manshu encountered a protest over the faculty-reservation dispute while traveling in India. All errors are our own.}
}
\author{ Manshu Khanna\thanks{Peking University HSBC Business School, Shenzhen 518055, China. Email: manshu@phbs.pku.edu.cn.} \and Haydar Evren\thanks{Akuna Capital, 24 School St, Boston, MA 02108, USA. Email: haydaremin@gmail.com.}}
\date{\today}
\usepackage[nameinlink]{cleveref} 

\begin{document}
\maketitle

\begin{abstract}
India's reservation policy entitles designated groups to fixed fractions of faculty positions. Positions are indivisible, arrive over time across departments, and appointments are irreversible. We show that no rule can keep cumulative reserved counts nondecreasing while rounding entitlements to adjacent integers at both the department and university levels; university-level compliance forces department-level errors that grow without bound. This impossibility explains why both solutions practiced in India provoked litigation and protests. A random-roster solution preserves irreversibility and department-level compliance, matches entitlements in expectation, and concentrates university-level errors. Dropping cumulative accounting restores exact rounding at both levels.
\end{abstract}
\vspace{.1in}
\noindent\textbf{Keywords:} Market Design, Indivisibility, Random Allocation, Intersecting Constraints, Apportionment

\newpage\doublespacing

\section{Introduction}\label{section:intro}
Market design has developed a rich toolkit for implementing fractional assignments with indivisible objects, including rounding schemes, decomposition theorems, and bounds on integer-allocation errors \citep{budish2013designing, akbarpour2020approximate}. What is often unclear is when these rounding constraints matter in an institution. In many static settings, capacities are fixed in advance, and it is hard to say in what sense a one-seat violation would even be observed, let alone litigated.

This paper studies an institution in which rounding is not an implementation detail but the substance of the policy conflict. India's reservation policy entitles Scheduled Castes (15\%), Scheduled Tribes (7.5\%), Other Backward Classes (27\%), and Economically Weaker Sections (10\%) to fixed fractions of public positions, and requires the resulting seat counts to be advertised, publicly auditable, and tracked {cumulatively} over recruitment cycles.\footnote{Cumulative tracking is operationalized through \emph{rosters}: running registers, maintained year over year, that attach each sanctioned post to a category (\Cref{section:shortcomings}). Universities publish these registers so that any citizen can verify why a reserved vacancy arose in one department and not another; \Cref{fig:exhibit-register} reproduces one.} A reservation policy specifies percentages, but a recruitment advertisement must specify whole numbers. When recruitment takes place across several departments, proportionality has two natural meanings: representation within each department and representation across the university. A procedure that protects one need not protect the other.

India's faculty-reservation dispute makes this tension concrete. Pooling vacancies across a university can protect aggregate representation while concentrating reserved recruitment in particular departments. Applying the same roster separately to small departments can instead produce an aggregate shortfall. All three branches of the Indian state have intervened in how to round these entitlements: the executive mandated pooling vacancies at the university level; the judiciary struck that down in favor of department-level rosters; nationwide protests then pushed Parliament to overturn the courts by statute (\Cref{section:institution}).

Our central result explains why no branch could succeed. Three individually natural requirements---\emph{monotonicity} (cumulative reserved counts never decrease), \emph{department quota compliance} (each department's cumulative allocation stays within one seat of its fractional entitlement), and \emph{university quota compliance} (the same at the university level)---are mutually incompatible (\Cref{cor:impossibility}). The deeper version of the result is starker: monotonicity together with exact university-level compliance forces department-level distortions that grow \emph{without bound} (\Cref{thm:impossibility}). The conflict between the Government's solution and the Court's solution cannot therefore be resolved merely by choosing a better ordering of departments on a common roster.

We propose a principled (second best) response. Preserving monotonicity and exact department-level compliance, we construct a \emph{random roster} solution (\Cref{thm:randomroster}): each department independently draws a roster from a distribution built by decomposing a bihierarchical constraint structure \citep{budish2013designing}. The solution is ex ante proportional everywhere. Because departments' draws are independent, university-level deviations concentrate: the probability that the university-level allocation of any category deviates from its entitlement by more than $b$ decays at Chernoff rates, at least at rate $e^{-b^2/(3|\mathcal{D})|}$ (\Cref{thm:approx}), placing the construction within the approximate-implementation paradigm of \citet{akbarpour2020approximate}. The solution thus retains department quota compliance while replacing exact aggregate quota compliance with ex ante proportionality and bounds on aggregate violations.

A static benchmark isolates the role of the cumulative requirement. Monotonicity, as practiced in India, bundles two distinct ideas: the irreversibility of an individual appointment, and the use of {perpetually cumulative} reservation counts that may never decrease. Only the second is essential to the impossibility. If the institution resets or rolls its accounting window, the period problem becomes static, and controlled rounding implements both quota levels exactly and proportionally (\Cref{thm:static}). \Cref{thm:static} thereby identifies a precise institutional reform that would dissolve the conflict; whether such a reform is politically or legally feasible in India is a separate question on which we take no position (\Cref{section:monotonicity}).

Although our formal analysis is disciplined by the Central Educational Institutions (Reservation in Teachers' Cadre) Act, 2019, the design pattern is not peculiar to Indian universities. Whenever a policy states entitlements as fractions, allocates indivisible units at two intersecting levels, and keeps a running account of past allocations, the same conflict appears. \Cref{section:pattern} documents instances ranging from office-wise recruitment at the Reserve Bank of India and cadre-wise allocation in the All India Services to district quotas in Sri Lankan university admissions and ethnic quotas in Singapore's public housing, and uses them to show which of our results applies to each.

\medskip
\noindent \textbf{Related literature.} 
The implementation of fractional assignments by lotteries over integral ones originates with \citet{hylland1979efficient}; \citet{bogomolnaia2001new} and successors study its fairness and incentive properties. \citet{budish2013designing} characterize when constraint structures (bihierarchies) admit exact decomposition, and \citet{akbarpour2020approximate} show how constraints outside a bihierarchy can be honored approximately, with high probability. Our constraint structure---monotonicity, department quotas, and university quotas jointly---is not a bihierarchy, and the university-level constraints cannot be placed at the deepest level of one, so neither result applies off the shelf.

Relative to the apportionment literature, the relationships are as follows. \emph{Biproportional apportionment} \citep{balinski1989axiomatic, pukelsheim2017proportional} rounds a two-way table subject to row and column totals, but is static: there is no cumulative constraint linking successive tables. \emph{Controlled rounding} \citep{cox1987constructive} solves exactly that static problem; we use it verbatim in \Cref{thm:static}, precisely to show that the static version of our problem is easy. \emph{House-monotone apportionment} \citep{balinski2010fair} and its stochastic version \citep{grimmett2004stochastic, golz2022apportionment} impose monotonicity as seats are added, but in one dimension (parties only); our monotonicity is its two-dimensional, cumulative cousin, and the interaction with a second (intersecting) quota dimension is what generates the impossibility. In one dimension, monotonicity and quota are compatible; it is the second dimension that breaks them.

To be explicit about what is new: the formalization of the cumulative implementation problem and the impossibility theorem (\Cref{thm:impossibility}) are new. The random-roster construction (\Cref{thm:randomroster,thm:approx}) is a new combination of existing tools---\citet{budish2013designing}'s decomposition applied to a roster constraint structure, with cross-department independence engineered so that Chernoff bounds \citep{chernoff1952measure} deliver the concentration property of \citet{akbarpour2020approximate}. \Cref{thm:static} is an application of \citet{cox1987constructive}. As a by-product, the random roster yields an alternative proof of the main possibility result of \citet{golz2022apportionment} (their Theorem 6).

Finally, a large market-design literature studies affirmative action given reserved seats \citep{abdulkadirouglu2003school, ehlers2014school, echenique2015control, Ayguen2017, dur2019explicit, sonmez2019affirmative, Aygun2021}. That literature takes the number of reserved seats as input; our paper is about how those numbers can coherently be produced. The random-roster solution we propose fits the minimalist market-design paradigm articulated in \citet{sonmez2025minimalist}: rather than overturning India’s cumulative roster institution, we diagnose its root failure and replace exact aggregate compliance with a statistical guarantee.

\section{Institutional Background} \label{section:institution}
The 1950 Constitution of India provides a basis for positive discrimination in favor of disadvantaged groups through \textit{reservation policies}. These policies mandate exclusive access to a fixed percentage of government jobs and seats in publicly funded institutions for members of Scheduled Castes (SC, 15\%), Scheduled Tribes (ST, 7.5\%), Other Backward Classes (OBC, 27\%), and Economically Weaker Sections (EWS, 10\%). For transparency, the number of reserved seats for each category is publicly advertised before every admission or recruitment cycle (\Cref{fig:exampleadv}).

Nowhere have the procedures for calculating reserved seats been more contentious than in universities. Eligibility and selection criteria for the same faculty rank differ across departments, so faculty vacancies are not interchangeable across a university. Each vacancy therefore belongs to two units at once, a department and the university, and the reservation policy speaks to both. This feature drew in all three arms of the Indian state.

\textbf{The Executive.} In August 2006, the University Grants Commission (UGC) issued \textit{Guidelines for Strict Implementation of Reservation Policy of the Government in Universities}, prohibiting the practice of treating the \textit{department as the unit} for calculating reserved seats and mandating the \textit{university as the unit}, with vacancies clubbed across departments within each rank (clauses 6(c) and 8(a)(v)).\footnote{The UGC is the statutory body responsible for implementing central government policy on admissions and recruitment in universities. Document last accessed on 22 February 2024 at \url{https://www.ugc.ac.in/pdfnews/7633178_English.pdf}.}

\textbf{The Judiciary.} In April 2017, the Allahabad High Court quashed those clauses and required the department to be the unit, arguing that the university as the unit ``would be not only impracticable, unworkable but also unfair and unreasonable'':\footnote{Judgement last accessed on 22 February 2024 at \url{https://indiankanoon.org/doc/177500970/}.}

\begin{quote}
If the University is taken as a `Unit' for every level of teaching and applying the roster, it could result in some departments/subjects having all reserved candidates and some having only unreserved candidates. Such a proposition again would be discriminatory and unreasonable.
\end{quote}

Following the court order, advertised reserved vacancies fell sharply. \Cref{table:exampletable} shows the case of Banaras Hindu University, where unreserved seats rose from 1188 under the quashed government solution to 1562 under the court's solution.\footnote{Last accessed on 22 February 2024 at \url{https://indianexpress.com/article/explained/hrd-ministry-ordinance-teacher-quota-university-prakash-javadekar-5616157/}.} The mechanics behind the collapse are simple. Reservations are administered through a \textit{roster}, a fixed sequence in which categories take turns claiming posts (\Cref{section:shortcomings}). Under the official rosters, the first post for SC arrives at roster point 7 and the first for ST only at point 14 (\Cref{fig:200pointroster}; both points are visible in the departmental register of \Cref{fig:exhibit-register}). Departments with a small number of posts cannot reserve seats for those categories, and most departments are small. The court's solution therefore starves the reserved categories at the university level, which sparked teachers' union protests across India.

\textbf{The Legislative.} The government appealed. ``How can the post of professor of Anatomy be compared with the professor of Geography? Are you clubbing oranges with apples?'' asked the Supreme Court, rejecting the appeal and calling the High Court judgment ``logical''.\footnote{Last accessed on 22 February 2024 at \url{https://main.sci.gov.in/supremecourt/2019/5495/5495_2019_Order_27-Feb-2019.pdf}.} Facing an aggrieved vote bank days before a general election, the government promulgated an ordinance in March 2019 restoring the university as the unit, now an Act of Parliament and the law in India.\footnote{Last accessed on 22 February 2024 at \url{http://egazette.gov.in/WriteReadData/2019/206575.pdf}.} The court's warning that university-level rosters ``could result in some departments/subjects having all reserved candidates and some having only unreserved candidates'' remains unanswered, and it motivates this paper.

\begin{table}[!h]
\centering
\ra{1.3}
\begin{adjustbox}{max width=1\textwidth}
\begin{tabular}{@{}lrrrrrcrrrrr@{}}\toprule
& \multicolumn{5}{c}{University as a Unit} & \phantom{abc}& \multicolumn{5}{c}{Department as a Unit} \\
& \multicolumn{5}{c}{(Government's Solution)} & \phantom{abc}& \multicolumn{5}{c}{(Court's Solution)} \\
\cmidrule{2-6} \cmidrule{8-12}
Position & General & SC & ST & OBC & Total && General & SC & ST & OBC & Total \\ \midrule
Professor & 197 & 38 & 18 & 0 & 253 && 250 & 3 & 0 & 0 & 253\\
Associate Professor& 410 & 79 & 39 & 0 & 528  && 500 & 25 & 3 & 0  & 528 \\
Assistant Professor& 581 & 172 & 86 & 310 & 1149  && 812 & 91 & 26 & 220  & 1149 \\
\midrule
Total& 1188 & 289 & 143 & 310 & 1930 && 1562 & 119 & 29 & 220  & 1930 \\
\bottomrule
\end{tabular}
\floatfoot{\emph{Notes:} Data shared in the government's Special Leave Petition filed in the Supreme Court of India. The advertisements underlying these data predate the operational rollout of the EWS quota in university teaching recruitment, so no EWS vacancies appear. OBC reservation did not apply to the two senior ranks at the time.}
\caption{NUMBER OF RESERVED VACANCIES IN BANARAS HINDU UNIVERSITY}
\label{table:exampletable}
\end{adjustbox}
\end{table}

Each intervention protected one margin by sacrificing the other. This paper explains why that pattern is not accidental. Under India's cumulative roster system, no allocation rule can guarantee both department-level and university-level proportionality for every possible sequence of vacancies. The next section formulates this design conflict.

\section{Formulation} \label{section:model}

A \textbf{problem of reservation in three dimensions} in period $t\in \mathbb{N}$ is a quadruple $\Lambda^t = ( \mathcal{D}, \mathcal{C} , \bm{\upalpha}, (\mathbf{q}^{s})_{s=1}^t )$. $\mathcal{D}$ and $\mathcal{C}$ are finite sets of \textbf{departments} and \textbf{categories} with $|\mathcal{D}| \geq 2$ and $|\mathcal{C}| \geq 2$. The \textbf{reservation scheme} is defined by a vector of fractions $\bm{\upalpha} = [\alpha_c]_{c\in \mathcal{C}}$. For each category $c \in \mathcal{C}$, $\alpha_c \in (0,1)$ is a rational number representing the fraction of vacancies to be reserved, with $\sum_{c \in \mathcal{C}} \alpha_c =1$. $\mathbf{q}^{s}= [q_d^{s}]_{d \in \mathcal{D}}$ is the \textbf{vector of vacancies} arriving in period $s \in \{1,2, \dotso, t \}$, where $q_d^s \in \mathbb{Z}_{\geq 0}$ is the number of vacancies arriving in department $d$.\footnote{A vacancy here is a newly sanctioned post. Under India's post-based rosters (\Cref{section:shortcomings}), a post that falls vacant through retirement or resignation is refilled in the category attached to its roster point, so replacement recruitment leaves the cumulative category counts unchanged and does not enter $\mathbf{q}^s$.} Let $Q_{d}^t :=\sum_{s \leq t} q_{d}^{s}$ denote the \textbf{period-$t$ (cumulative) sum of vacancies in department $d$}.

\textbf{Tables.} Throughout, allocation problems are summarized by two-way tables. Rows are departments plus a final \emph{university-total} row (indexed $\uni$); columns are beneficiary categories plus a final \emph{department-size} column (indexed $\tot$). Internal entries are department-by-category assignments; the university-total row collects column sums; the department-size column collects row sums. Every assignment table in the paper has this shape, and every property we study constrains either the internal entries (local proportionality), the university row (aggregate proportionality), or the table's evolution over time (monotonicity).

A \textbf{period-$t$ (cumulative) fractional assignment table} for problem $\Lambda^t$ is a two-way table $X^t$ with rows indexed by $\mathcal{D} \cup \{\uni\}$ and columns by $\mathcal{C} \cup \{\tot\}$, such that internal entries $x_{d,c}^t = \alpha_{c} Q_{d}^t$ for any $d \in \mathcal{D}$ and $c \in \mathcal{C}$, row total entries $x_{d,\tot}^t=Q_{d}^t$, column total entries $x_{\uni,c}^t=\alpha_{c} \sum_{d \in \mathcal{D}} Q_{d}^t$, and grand total entry $x_{\uni,\tot}^t=\sum_{d \in \mathcal{D}} Q_{d}^t$. The internal entry $x_{d,c}^t$ is the \textbf{period-$t$ (cumulative) fractional assignment for category $c$ in department $d$}; the column total $x_{\uni,c}^t$ is the corresponding assignment \textbf{in the university}. We denote the set of fractional assignment tables by $\mathcal{X}$.

For instance, consider the problem $\Lambda^2 = ( \{d_1,\; d_2 \}, \{c_1, \; c_2 \} , \bm{\upalpha}=[0.1 , 0.9], (\mathbf{q}^{1},\mathbf{q}^{2}) = ([9,8],[17,7]) )$. \Cref{fig: fairtables} illustrates its period-1 and period-2 fractional assignment tables. The reservation scheme reserves 10\% of vacancies in the university for members of category $c_1$. In period-1, departments $d_1$ and $d_2$ have 9 and 8 vacancies; in period-2, they receive 17 and 7 more, so period-2 cumulative sums are 26 and 15, shown in the size column.

\begin{figure}[!ht]
  \centering
  \begin{subfigure}{.475\linewidth}
        \centering
        \textbf{$X^1$} =
        \begin{tabular}{cc|c}
            0.9 & 8.1 & 9\tabularnewline
            0.8 & 7.2 & 8\tabularnewline
            \hline 
            1.7 & 15.3 & 17\tabularnewline
        \end{tabular}
    \caption{PERIOD-1 FRACTIONAL ASSIGNMENT}
  \end{subfigure}%
  \hspace{0.1em}%
  \begin{subfigure}{.475\linewidth}
        \centering
        \textbf{$X^2$} =
        \begin{tabular}{cc|c}
            2.6 & 23.4 & 26\tabularnewline
            1.5 & 13.5 & 15\tabularnewline
            \hline 
            4.1 & 36.9 & 41\tabularnewline
        \end{tabular}
            \caption{PERIOD-2 FRACTIONAL ASSIGNMENT}
  \end{subfigure}%
      \caption{FRACTIONAL ASSIGNMENT TABLES}
  \label{fig: fairtables}
\end{figure}

A \textbf{period-$t$ sequence of (cumulative) fractional assignment tables} for the problem $\Lambda^t$ is a sequence $Y^t=(X^{1},\dotso,X^t)$, where $X^s$ is the period-$s$ fractional assignment table for any $s \in \{1,2,\dotso,t \}$. By $\mathcal{Y}^t$ we denote the set of all period-$t$ sequences. Given $Y^t$, if $Y^t=(Y^{t-1},X^{t})$ for some $X^{t} \in \mathcal{X}$, we say that $Y^t$ \textbf{follows} $Y^{t-1}$. We index a solution by the full history $Y^t$ rather than by the current cumulative table $X^t$ alone because monotonicity constrains the period-$t$ integral assignment relative to assignments made in earlier periods. Two histories leading to the same $X^t$ can therefore induce different past integral assignments and hence different feasible sets in period $t$.

A two-way table is \textbf{additive} if entries add along the rows and columns to all corresponding totals. A \textbf{period-$t$ (cumulative) integral assignment table} for problem $\Lambda^t$ is a non-negative integer two-way table $\Bar{X}^t= (\Bar{x}_{d,c}^t)$ of the same shape, such that $\Bar{X}^t$ is additive and $\Bar{x}_{d,\tot}^{t}=x_{d,\tot}^{t}$ for any $d \in \mathcal{D}$. We denote the set of integral assignment tables by $\Bar{\mathcal{X}}$.

\subsection{Deterministic solutions and properties}

A \textbf{deterministic solution} $R:\cup_{s \in \mathbb{N}}\mathcal{Y}^s \rightarrow \Bar{\mathcal{X}}$ maps each sequence of fractional assignment tables to an integral assignment table. We will use $R(y_{d,c}^t)$ to denote the entry $\bar{x}^t_{d,c}$ of $R(Y^t)$.

The \textbf{bias of a deterministic solution} $R$ at $Y^t$ is a two-way table $\bias(R(Y^t))$, with each entry $\bias(R(y_{d,c}^t)) := R(y_{d,c}^t) -x_{d,c}^t$: the difference between the solution and the fractional assignment table.

\begin{itemize}
    \item A deterministic solution $R$ is \textbf{monotonic} if $R(Y^{t}) \geq R(Y^{t-1})$ for any $Y^t \in \cup_{s\in \mathbb{N}}\mathcal{Y}^s$ that follows $Y^{t-1}$.\footnote{The relation ``$\geq$'' compares tables entry-wise.} Going from period $t$ to period $t+1$, the cumulative integral assignments of each category can only weakly increase: reservations are irreversible. Monotonicity is essential under India's current accounting system; \Cref{section:monotonicity} examines its content.

    \item A deterministic solution $R$ is \textbf{locally proportional} if $|\bias(R(y_{d,c}^t)) |< 1$ for each \textit{internal} entry ($d \in \mathcal{D}$, $c \in \mathcal{C}$) for any $Y^t \in \cup_{s\in \mathbb{N}}\mathcal{Y}^s$: each category's cumulative fractional assignment in each department is rounded to an adjacent integer.\footnote{For any $x \in \mathds{R}$, $\floor*{x}$ is the floor of $x$ and $\ceil*{x}$ the ceiling of $x$.} This is essentially the ideal integer solution that can be achieved at the department level.

    \item A deterministic solution $R$ is \textbf{aggregately proportional} if $|\bias(R(y_{\uni,c}^t)) |< 1$ for each \textit{column total} entry ($c \in \mathcal{C}$) for any $Y^t \in \cup_{s\in \mathbb{N}}\mathcal{Y}^s$: the same idea at the university level.

    \item A deterministic solution $R$ has a \textbf{finite bias} if there exists a constant $b >0$ such that $|\bias(R(y_{d,c}^t))| < b$ for each entry and any $Y^t \in \cup_{s\in \mathbb{N}}\mathcal{Y}^s$. This relaxes the previous two properties combined by requiring only finite deviations.
\end{itemize}

Reconsider the problem in \Cref{fig: fairtables}. \Cref{fig: restables} illustrates two possible monotonic deterministic solutions: $R_1$ is both locally and aggregately proportional at the displayed histories, while $R_2$ is locally proportional only.

\begin{figure}[!htb]
    \centering
    \begin{subfigure}[b]{0.475\textwidth}
        \centering
        \textbf{$X^1$} =
        \begin{tabular}{cc|c}
            0.9 & 8.1 & 9\tabularnewline
            0.8 & 7.2 & 8\tabularnewline
            \hline 
            1.7 & 15.3 & 17\tabularnewline
        \end{tabular}
        \caption{PERIOD-1  FRACTIONAL ASSIGNMENT}
    \end{subfigure}
    \hfill
    \begin{subfigure}[b]{0.475\textwidth}  
       \centering
        \textbf{$X^2$} =
        \begin{tabular}{cc|c}
            2.6 & 23.4 & 26\tabularnewline
            1.5 & 13.5 & 15\tabularnewline
            \hline 
            4.1 & 36.9 & 41\tabularnewline
        \end{tabular}
        \caption{PERIOD-2  FRACTIONAL ASSIGNMENT}
    \end{subfigure}
    \vskip\baselineskip
    \begin{subfigure}[b]{0.475\textwidth}
        \centering
        \textbf{$R_1(Y^1)$} =
        \begin{tabular}{cc|c}
            1 & 8 & 9\tabularnewline
            1 & 7 & 8\tabularnewline
            \hline 
            2 & 15 & 17\tabularnewline
        \end{tabular}
        \caption{PERIOD-1 INTEGRAL ASSIGNMENT}
    \end{subfigure}
    \hfill
    \begin{subfigure}[b]{0.475\textwidth}  
       \centering
        \textbf{$R_1(Y^2)$} =
        \begin{tabular}{cc|c}
            3 & 23 & 26\tabularnewline
            1 & 14 & 15\tabularnewline
            \hline 
            4 & 37 & 41\tabularnewline
        \end{tabular}
        \caption{PERIOD-2 INTEGRAL ASSIGNMENT}
    \end{subfigure}
        \vskip\baselineskip
    \begin{subfigure}[b]{0.475\textwidth}
        \centering
        \textbf{$R_2(Y^1)$} =
        \begin{tabular}{cc|c}
            0 & 9 & 9\tabularnewline
            0 & 8 & 8\tabularnewline
            \hline 
            0 & 17 & 17\tabularnewline
        \end{tabular}
        \caption{PERIOD-1 INTEGRAL ASSIGNMENT}
    \end{subfigure}
    \hfill
    \begin{subfigure}[b]{0.475\textwidth}  
       \centering
        \textbf{$R_2(Y^2)$} =
        \begin{tabular}{cc|c}
            3 & 23 & 26\tabularnewline
            1 & 14 & 15\tabularnewline
            \hline 
            4 & 37 & 41\tabularnewline
        \end{tabular}
        \caption{PERIOD-2 INTEGRAL ASSIGNMENT}
    \end{subfigure}
  \caption{TWO DETERMINISTIC SOLUTIONS}
  \label{fig: restables}
\end{figure}

\subsection{Randomized solutions and properties}

A \textbf{randomized solution} is a probability distribution $\phi$ over a finite set of deterministic solutions, where $\phi(R)$ denotes the probability of deterministic solution $R$. Let $\tilde{R} \sim \phi$ denote the (random) realized rule, and for any $Y^t \in \cup_{s\in \mathbb{N}}\mathcal{Y}^s$ let $Z^t := \tilde{R}(Y^t)$ be the resulting random integral assignment table, with entries $z_{d,c}^t$. One can view $\phi$ as a procedure initialized with a randomly chosen seed that takes $Y^t$ as input and returns an integral assignment.\footnote{This is akin to the definition of randomized apportionment methods, where the randomness of the methods is determined by an implicit random seed (\cite{golz2022apportionment}).}

\begin{itemize}
\item A randomized solution $\phi$ is \textbf{monotonic}, \textbf{locally proportional}, or \textbf{aggregately proportional} if all deterministic solutions in its support satisfy the respective property.

\item A randomized solution $\phi$ is \textbf{ex ante proportional} if, for any $Y^t \in \cup_{s\in \mathbb{N}}\mathcal{Y}^s$,
$$ \mathbb{E}_{\phi}\big[\tilde{R}(Y^t)\big] \;=\; \sum_{R} \phi(R)\, R(Y^t) \;=\; X^t .$$

\item A randomized solution $\phi$ \textbf{satisfies approximate aggregate proportionality} if, for any $Y^t \in \cup_{s\in \mathbb{N}}\mathcal{Y}^s$, any category $c \in \mathcal{C}$, and any $b >0$,
$$\text{Pr}\big(z_{\uni,c}^t-x_{\uni,c}^t \geq b\big) < e^{-\frac{b^2}{3|\mathcal{D}|} }\text{ ,}
\qquad
\text{Pr}\big(z_{\uni,c}^t-x_{\uni,c}^t \leq -b\big) < e^{-\frac{b^2}{2|\mathcal{D}|} }\text{ .}$$
These are Chernoff-type concentration bounds (\cite{chernoff1952measure}): the probability that the university-level allocation deviates from its fractional assignment by more than $b$ decays exponentially with $b^2$.
\end{itemize}

\section{Rosters and the two solutions practiced in India}\label{section:shortcomings}

Reservations in India are administered through rosters. Formally, a \textbf{roster} $\sigma: \{1,2, \dotso \} \rightarrow \mathcal{C}$ is an ordered list over the set of categories $\mathcal{C}$, assigning each successive post to a category. The Government of India prescribes a 13-point roster for smaller cadres and a 200-point roster for cadres with fourteen or more posts (\Cref{fig:13pointroster,fig:200pointroster}).\footnote{\citet{KhannaEvren2025} analyze the construction of these rosters as a sequence of connected apportionment problems and show that the Webster--Sainte-Lagu\"e method is the unique divisor method satisfying natural fairness and practicality criteria.} Two features of how rosters are operated matter for everything that follows.

\emph{Rosters are running accounts.} Departmental rules require that ``the roster register shall be maintained in the form of a running account year after year. For example if recruitment in a year stops at point 6, recruitment in the following year shall begin from point 7,'' and that ``fresh cycle of roster shall be started after completion of all the points in the roster.''\footnote{See page 1 of \url{https://dopt.gov.in/sites/default/files/Ch-05_2014.pdf}, last accessed on 22 February 2024.} A publicly declared roster of length $k$ therefore pins down the allocation of an infinite sequence of posts as a concatenation of blocks of length $k$: the categories of the first $q^1$ posts are $\sigma(1),\dotso, \sigma(q^1)$; of the next $q^2$ posts, $\sigma(q^1 +1),\dotso, \sigma(q^1 + q^2)$; and so forth.

\emph{Rosters are post-based.} A roster is a template for taking turns, not a hiring target. A department with six sanctioned posts occupies points 1--6 of the roster and nothing more; point 7 comes into play only if a seventh post is created. Each point is permanently attached to a post. When the incumbent retires or resigns, the post is refilled against the same point in the same category, so the number of posts reserved for a category in a department can never fall.\footnote{This is the ``post-based'' roster system adopted after \emph{R.K.\ Sabharwal v.\ State of Punjab} (1995), replacing earlier ``vacancy-based'' rosters. Under a vacancy-based roster every recruitment, including a replacement, advances the roster; under a post-based roster only new posts do.} In the notation of \Cref{section:model}, $q_d^s$ counts new posts, and monotonicity is literally the rule that a point, once designated, is never re-designated. \Cref{fig:exhibit-register} reproduces a running register of exactly this kind, maintained by Hindu College, University of Delhi, for its Assistant Professor cadre. For each department it lists every roster point, the category attached to it, the appointee and date of appointment, the category in which the post was actually filled, and any excess or shortfall carried forward. In the Botany department, points 1--3 are unreserved and were filled in 1988, 1991, and 1995; the department's first reserved point is point 4 (OBC), and its first SC point is point 7. Any citizen can open this register and verify why a reserved post arose in Botany and not in Economics.

Maintaining rosters is central to the implementation of reservations in India, but for universities it raises the question that generated the dispute: does each department maintain its own roster, or does the university maintain one? The \emph{Government's solution} $R_G$ treats the \textit{university as the unit}: posts across all departments are pooled, and one roster is maintained at the university level.\footnote{When pooling vacancies across departments, a fixed order over departments is required; in India, the alphabetic order is used.} The \emph{Court's solution} $R_C$ treats the \textit{department as the unit}: each department independently maintains the roster, as in \Cref{fig:exhibit-register}.

\Cref{example:ex1} exhibits both failures at once; \Cref{fig:court_gov} displays the outcomes, and the period-by-period computation is in \Cref{appendix:indiaexamples}.

\begin{example}\label{example:ex1}
Consider a problem $\Lambda^3 = ( \{d_1,\; d_2,\; d_3 ,\; d_4 \}, \{c_1, \; c_2 \} , \bm{\upalpha}=[1/3 , 2/3], (\mathbf{q}^{1},\mathbf{q}^{2},\mathbf{q}^{3}) = ([2,1,2,1],[2,1,2,1], [2,1,2,1]) )$. The roster is
\[
    \sigma(k)= 
\begin{dcases}
    c_1,& \text{if } k \text{ is a multiple of } 3 \\
    c_2,              & \text{otherwise.}
\end{dcases}
\]
\end{example}

\begin{figure}[!htb]
    \centering
    \begin{subfigure}{.33\linewidth}
        \centering
        \textbf{$X^1$} =
        \begin{tabular}{cc|c}
            2/3 & 4/3 & 2\tabularnewline
            1/3 & 2/3 & 1\tabularnewline
            2/3 & 4/3 & 2\tabularnewline
            1/3 & 2/3 & 1\tabularnewline
            \hline 
            2 & 4 & 6\tabularnewline
        \end{tabular}
            \caption{\centering{PERIOD-1 FRACTIONAL ASSIGNMENT}}
  \end{subfigure}%
  \hspace{0.1em}%
  \begin{subfigure}{.33\linewidth}
        \centering
        \textbf{$X^2$} =
        \begin{tabular}{cc|c}
            4/3 & 8/3 & 4\tabularnewline
            2/3 & 4/3 & 2\tabularnewline
            4/3 & 8/3 & 4\tabularnewline
            2/3 & 4/3 & 2\tabularnewline
            \hline 
            4 & 8 & 12\tabularnewline
        \end{tabular}
            \caption{\centering{PERIOD-2 FRACTIONAL ASSIGNMENT}}
  \end{subfigure}%
  \hspace{0.1em}%
  \begin{subfigure}{.33\linewidth}
        \centering
        \textbf{$X^3$} =
        \begin{tabular}{cc|c}
            2 & 4 & 6\tabularnewline
            1 & 2 & 3\tabularnewline
            2 & 4 & 6\tabularnewline
            1 & 2 & 3\tabularnewline
            \hline 
            6 & 12 & 18\tabularnewline
        \end{tabular}
            \caption{\centering{PERIOD-3 FRACTIONAL ASSIGNMENT}}
  \end{subfigure}
    
    \vskip\baselineskip
    \begin{subfigure}{.33\linewidth}
        \centering
        \textbf{$R_{G}(Y^1)$} =
        \begin{tabular}{cc|c}
            0 & 2 & 2\tabularnewline
            1 & 0 & 1\tabularnewline
            0 & 2 & 2\tabularnewline
            1 & 0 & 1\tabularnewline
            \hline 
            2 & 4 & 6\tabularnewline
        \end{tabular}
   \caption{\centering{PERIOD-1 INTEGRAL ASSIGNMENT}}
  \end{subfigure}%
  \hspace{0.1em}%
  \begin{subfigure}{.33\linewidth}
        \centering
        \textbf{$R_{G}(Y^2)$} =
        \begin{tabular}{cc|c}
            0 & 4 & 4\tabularnewline
            2 & 0 & 2\tabularnewline
            0 & 4 & 4\tabularnewline
            2 & 0 & 2\tabularnewline
            \hline 
            4 & 8 & 12\tabularnewline
        \end{tabular}
   \caption{\centering{PERIOD-2 INTEGRAL ASSIGNMENT}}
  \end{subfigure}%
  \hspace{0.1em}%
  \begin{subfigure}{.33\linewidth}
        \centering
        \textbf{$R_{G}(Y^3)$} =
        \begin{tabular}{cc|c}
            0 & 6 & 6\tabularnewline
            3 & 0 & 3\tabularnewline
            0 & 6 & 6\tabularnewline
            3 & 0 & 3\tabularnewline
            \hline 
            6 & 12 & 18\tabularnewline
        \end{tabular}
   \caption{\centering{PERIOD-3 INTEGRAL ASSIGNMENT}}
  \end{subfigure}
    
        \vskip\baselineskip
    \begin{subfigure}{.33\linewidth}
        \centering
        \textbf{$R_{C}(Y^1)$} =
        \begin{tabular}{cc|c}
            0 & 2 & 2\tabularnewline
            0 & 1 & 1\tabularnewline
            0 & 2 & 2\tabularnewline
            0 & 1 & 1\tabularnewline
            \hline 
            0 & 6 & 6\tabularnewline
        \end{tabular}
   \caption{\centering{PERIOD-1 INTEGRAL ASSIGNMENT}}
  \end{subfigure}%
  \hspace{0.1em}%
  \begin{subfigure}{.33\linewidth}
        \centering
        \textbf{$R_{C}(Y^2)$} =
        \begin{tabular}{cc|c}
            1 & 3 & 4\tabularnewline
            0 & 2 & 2\tabularnewline
            1 & 3 & 4\tabularnewline
            0 & 2 & 2\tabularnewline
            \hline 
            2 & 10 & 12\tabularnewline
        \end{tabular}
   \caption{\centering{PERIOD-2 INTEGRAL ASSIGNMENT}}
  \end{subfigure}%
  \hspace{0.1em}%
  \begin{subfigure}{.33\linewidth}
        \centering
        \textbf{$R_{C}(Y^3)$} =
        \begin{tabular}{cc|c}
            2 & 4 & 6\tabularnewline
            1 & 2 & 3\tabularnewline
            2 & 4 & 6\tabularnewline
            1 & 2 & 3\tabularnewline
            \hline 
            6 & 12 & 18\tabularnewline
        \end{tabular}
   \caption{\centering{PERIOD-3 INTEGRAL ASSIGNMENT}}
  \end{subfigure}
 \caption{COURT'S AND GOVERNMENT'S SOLUTION}
  \label{fig:court_gov}
\end{figure}

The fractional tables in \Cref{fig:court_gov} are obtained by multiplying each department's cumulative vacancies by the reservation shares $(1/3,2/3)$. The integral tables instead assign successive vacancies according to the repeating roster $(c_2,c_2,c_1)$. Under the Government's solution, period-1 vacancies are pooled and ordered by department: roster points 1--2 are assigned to $d_1$, point 3 to $d_2$, points 4--5 to $d_3$, and point 6 to $d_4$. Thus, the two $c_1$ points fall in $d_2$ and $d_4$. Because the same six-vacancy pattern recurs, this alignment repeats in every period. By period-3, $R_G$ therefore reserves 0 seats for $c_1$ in $d_1$ and $d_3$ (fractional assignment: 2 each) and 3 seats in $d_2$ and $d_4$ (fractional assignment: 1 each). Hence, $R_G$ is not locally proportional, and these department-level deviations grow without bound if the vacancy pattern persists. Under the Court's solution, each department starts its own copy of the roster. In period-1, $d_1$ and $d_3$ use roster points 1--2, while $d_2$ and $d_4$ use point 1; all these points belong to $c_2$, so no department reserves for $c_1$, compared with a university fractional assignment of 2. By period-2, $d_1$ and $d_3$ have each passed point 3 and reserved one seat for $c_1$, whereas $d_2$ and $d_4$ have reached only point 2 and reserved none. The university therefore reserves only 2 seats for $c_1$, compared with a fractional assignment of 4. Thus, $R_C$ is not aggregately proportional, and replicating the departments scales up the aggregate shortfall. This is precisely the mechanism behind the Banaras Hindu University figures in \Cref{table:exampletable}.

The choice of roster in \Cref{example:ex1} is not the source of these shortcomings. The source is that each solution operates at one level, either the department or the university, while the reservation policy binds at both. \Cref{thm:impossibility} shows that this is not a defect of rosters but of any monotonic accounting system.

\section{Results} \label{section:results}

\subsection{The impossibility}
Our first result shows that a monotonic solution cannot be both locally and aggregately proportional. In fact, a monotonic and aggregately proportional solution cannot keep department-level deviations bounded by \emph{any} finite number.

\begin{theorem}\label{thm:impossibility}
There does not exist a deterministic solution $R$ that is monotonic, aggregately proportional, and has a finite bias.
\end{theorem}

\begin{corollary}\label{cor:impossibility}
There does not exist a deterministic solution $R$ that is monotonic, locally proportional, and aggregately proportional.
\end{corollary}

Fix any deterministic solution $R$ that is monotonic and aggregately proportional. We construct an adaptive vacancy sequence along which $R$ accumulates unbounded department-level bias.

\begin{example} \label{example:multi prop2}
Consider a problem with three departments $d_1, d_2, d_3$, two categories $c_1,c_2$, and reservation scheme $\bm{\upalpha} = [0.5 , 0.5]$. The departments have $\mathbf{q}^1=[0,0,1]$ vacancies in period-1 and $\mathbf{q}^2=[1,0,0]$ in period-2.

Aggregate proportionality is equivalent to reserving exactly $k$ vacancies for each of $c_1$ and $c_2$ in every $2k$ cumulative vacancies in the university. In period-1, department $d_3$ can reserve its vacancy for either category; without loss of generality, say $c_1$. In period-2, with 2 cumulative vacancies in the university, exactly 1 must be reserved for each category, so department $d_1$ \emph{must} reserve its vacancy for $c_2$ (\Cref{fig:multi prop 2}).

\begin{figure}[!htb]
    \centering
    \begin{subfigure}[b]{0.475\textwidth}
        \centering
        \textbf{$X^1$} =
        \begin{tabular}{cc|c}
            0 & 0 & 0\tabularnewline
            0 & 0 & 0\tabularnewline
            0.5 & 0.5 & 1\tabularnewline
            \hline 
            0.5 & 0.5 & 1\tabularnewline
        \end{tabular}
        \caption{PERIOD-1  FRACTIONAL ASSIGNMENT}
    \end{subfigure}%
    \hspace{0.1em}
    \begin{subfigure}[b]{0.475\textwidth}
        \centering
        \textbf{$X^2$} =
        \begin{tabular}{cc|c}
            0.5 & 0.5 & 1\tabularnewline
            0 & 0 & 0\tabularnewline
            0.5 & 0.5 & 1\tabularnewline
            \hline 
            1 & 1 & 2\tabularnewline
        \end{tabular}
        \caption{PERIOD-2 FRACTIONAL ASSIGNMENT}
    \end{subfigure}
    
    \vskip\baselineskip
    \begin{subfigure}[b]{0.475\textwidth}
        \centering
        \textbf{$R(Y^1)$} =
        \begin{tabular}{cc|c}
            0 & 0 & 0\tabularnewline
            0 & 0 & 0\tabularnewline
            1 & 0 & 1\tabularnewline
            \hline 
            1 & 0 & 1\tabularnewline
        \end{tabular}
        \caption{PERIOD-1  INTEGRAL ASSIGNMENT}
    \end{subfigure}%
    \hspace{0.1em}
    \begin{subfigure}[b]{0.475\textwidth}  
       \centering
        \textbf{$R(Y^2)$} =
        \begin{tabular}{cc|c}
            0 & 1 & 1\tabularnewline
            0 & 0 & 0\tabularnewline
            1 & 0 & 1\tabularnewline
            \hline 
            1 & 1 & 2\tabularnewline
        \end{tabular}
        \caption{PERIOD-2  INTEGRAL ASSIGNMENT}
    \end{subfigure}
    
  \caption{PERIOD-1 AND PERIOD-2  ASSIGNMENT TABLES}
  \label{fig:multi prop 2}
\end{figure}

The construction now proceeds recursively. At the end of every even period $2r$, the university has $2r$ cumulative vacancies. Aggregate proportionality therefore requires exactly $r$ seats for each category. In period $2r+1$, the adversary gives the only new vacancy to $d_3$. Because the total number of vacancies rises by one, monotonicity implies that $R$ must add this vacancy to exactly one of the two categories. The adversary then chooses the location of the period-$(2r+2)$ vacancy after observing that assignment:

\textbf{Case A:} If $d_3$'s new vacancy is assigned to $c_1$, the adversary gives the next vacancy to $d_1$. Before that vacancy is assigned, the university has $r+1$ seats for $c_1$ and $r$ for $c_2$. At the end of period $2r+2$, aggregate proportionality requires $r+1$ seats for each category. Since monotonicity rules out reversing any earlier assignment, $d_1$'s new vacancy must be assigned to $c_2$.

\textbf{Case B:} If $d_3$'s new vacancy is assigned to $c_2$, the adversary instead gives the next vacancy to $d_2$. The symmetric argument forces this vacancy to be assigned to $c_1$.

The end of period $2r+2$ again has equal category totals, so the same construction can be repeated after every even period. It therefore defines an infinite vacancy history for any deterministic solution $R$. Among the infinitely many odd-period vacancies assigned in $d_3$, at least one category must be selected infinitely often. If $c_1$ is selected infinitely often, $d_1$ receives infinitely many vacancies, all assigned to $c_2$; after $N$ such vacancies, $R(y^t_{d_1,c_1})=0$ while $x^t_{d_1,c_1}=N/2$, giving absolute bias $N/2$. If $c_2$ is selected infinitely often, the symmetric conclusion holds for the $(d_2,c_2)$ entry. Thus, for any constant $b>0$, one of these entries has absolute bias greater than $b$ after finitely many periods. The complete table sequence through period-5 is displayed in \Cref{appendix:impossibilityexample}.
\end{example}

Because the construction applies to every deterministic $R$ that is monotonic and aggregately proportional, it proves \Cref{thm:impossibility}.\footnote{An example for any number of categories and departments can be constructed similarly. \Cref{example:multi prop2} is constructed so that it not only illustrates the failure but also demonstrates that every solution fails on some entry.} \Cref{cor:impossibility} follows because local proportionality implies finite bias: internal deviations are below one, and summing them across the finitely many departments bounds every university-level deviation.

\begin{remark}\label{remark:localside}
Although every locally proportional solution has finite aggregate bias, that bias need not be small. \Cref{example:100dept} in \Cref{appendix:100dept} shows monotonicity generating a university-level deviation of 25 seats while every department remains locally proportional.\footnote{For any category $c$ with $\alpha_c=\nicefrac{r}{w}$ in lowest terms, the absolute university-level deviation under local proportionality is bounded above by $|\mathcal D|\bigl(1-\nicefrac{1}{w}\bigr)$, and this bound can be attained. For example, with 100 departments of 3 vacancies each and $\alpha_c=\nicefrac{7}{20}$, each department's entitlement is $1.05$; a locally proportional assignment may allocate 2 seats per department to $c$, yielding 200 seats against a university-level entitlement of 105, an aggregate deviation of $100(1-\nicefrac{1}{20})=95$.}
\end{remark}

\subsection{What monotonicity does and does not mean}\label{section:monotonicity}
Monotonicity drives \Cref{thm:impossibility}, but three ideas should be distinguished:
\begin{enumerate}[(i)]
    \item the irreversibility of an individual appointment: a person hired cannot be un-hired;
    \item the use of \emph{perpetually cumulative} reservation counts that can never decrease, as under India's post-based running rosters; and
    \item accounting conventions that preserve (i) while relaxing (ii), such as \emph{attrition-based} accounting, under which counts decrease when faculty retire or resign, or \emph{rolling-window} accounting, under which only the last $H$ hires or years count toward compliance.
\end{enumerate}
Our monotonicity property formalizes (ii), as required by current Indian practice: rosters are maintained as running accounts year after year, and a point, once attached to a post, is never re-designated. But (ii) neither follows from (i) nor is inherent in affirmative action or proportional allocation. An institution could modify its accounting baseline without reversing an individual appointment. Attrition-based accounting, in particular, is exactly what the post-based roster rules out: when an incumbent retires, the post keeps its point's category rather than returning to the pool (\Cref{section:shortcomings}).

The impossibility arises because perpetual cumulative accounting allows early integer errors to constrain the feasible set forever. The alternatives in (iii) break this accumulation logic, while \Cref{thm:static} shows that abandoning cumulative accounting altogether restores exact implementability at both levels.

\subsection{Scope of the impossibility}\label{section:scope}

The vacancy sequence in the proof of \Cref{thm:impossibility} responds to the solution's previous assignments. The theorem is therefore a worst-case result over possible histories, including adversarially and adaptively selected vacancy sequences. This establishes the failure of a guarantee that must hold for every possible history; it does not claim that such histories are typical. When vacancy arrivals are random and independent over time, the adaptive construction is no longer available. Whether the three properties can then be achieved approximately or asymptotically requires a separate stochastic analysis.

The use of full histories is connected to this scope. Although $X^t$ records current cumulative fractional entitlements, monotonicity also constrains $R(Y^t)$ through the integral assignments made at earlier prefixes of $Y^t$. Two histories ending at the same $X^t$ can therefore have different past integral assignments and different feasible assignments in period $t$. This is why solutions are indexed by $Y^t$ rather than by $X^t$ alone; without monotonicity, the distinction disappears (\Cref{section:static}).

The construction also rules out more than a constant bias bound. At the end of period $t=2r$, department $d_3$ has received $r$ vacancies. At least $r/2$ of them must have been assigned to the same category. If that category is $c_1$, then $d_1$ has received at least $r/2$ vacancies, all forced to $c_2$, so the absolute bias in entry $(d_1,c_1)$ is at least $(r/2)(1/2)=r/4=t/8$. If the category is $c_2$, the same bound holds for entry $(d_2,c_2)$. Hence no bias bound $b^t=o(t)$ can be guaranteed.

\subsection{A principled second best}\label{section:secondbest}

Monotonicity is fixed by current institutional practice, and local proportionality is the margin whose violation the Allahabad High Court held unconstitutional. We therefore keep both exactly. Aggregate proportionality is then treated as a \emph{soft} constraint, a goal rather than a hard constraint, in the sense of \citet{akbarpour2020approximate}. This is a common response to constraint structures that cannot be satisfied jointly; see, e.g., \citet{budish2011combinatorial} on flexible course capacities and \citet{nguyen2018near} on near-feasible stable matchings.

\begin{theorem}\label{thm:randomroster}
There exists a randomized solution $\phi$ that is ex ante proportional, monotonic, and locally proportional.
\end{theorem}

Because future vacancy sequences are unknown, the solution must work after every possible sequence of cumulative fractional assignments. A roster is a natural way to achieve this consistency: once a department is assigned a roster, the category of every successive vacancy is determined, regardless of when those vacancies arrive. Local proportionality then requires every prefix of the roster to keep each category's cumulative allocation within one seat of its fractional entitlement. Since many rosters may satisfy these requirements, the construction randomizes among them so that the fractional entitlements are also recovered in expectation.

A complete proof is in \Cref{appendix:proofthm2}. The construction, which we call the \emph{random roster}, has two steps. First, represent the reservation scheme as a $k \times |\mathcal{C}|$ table, where $k$ is the least common multiple of the denominators in $\bm{\upalpha}$, rows are successive seat indices, and every entry in column $c$ equals $\alpha_c$. Requiring each row to assign one seat and each category's cumulative column sum to round its fractional value produces a bihierarchical constraint structure. The decomposition theorem of \cite{budish2013designing} therefore expresses the fractional table as a lottery over integral tables, each corresponding to a locally proportional roster. This decomposition can be computed in polynomial time using a network-flow algorithm (\Cref{Appendix: A.2,Appendix: A.2.2}). Second, independently draw one such roster for each department. Every realization is locally proportional and monotonic, while the lottery reproduces each department's fractional assignment in expectation, establishing ex ante proportionality. Operationally, nothing changes for a department except that the roster it follows is drawn by lot at the outset rather than copied from the government template; the register of \Cref{fig:exhibit-register} would be kept exactly as it is today.

\begin{remark}\label{remark:universityversion}
The same construction can instead assign a roster drawn from the lottery to the university as a whole. This yields a randomized solution that is ex ante proportional, monotonic, and aggregately proportional, although it need not be locally proportional.
\end{remark}

\Cref{thm:randomroster} retains monotonicity and exact local proportionality, but \Cref{thm:impossibility} rules out adding exact aggregate proportionality as another hard constraint. We therefore return to aggregate proportionality as a soft constraint: can the university-level allocation remain close to its fractional assignment with high probability?

This conclusion does not follow directly from the existing approximate-implementation results. Combining monotonicity with both local and aggregate proportionality destroys the bihierarchical structure used above. \cite{akbarpour2020approximate} show that soft constraints can be approximately implemented when they lie at the deepest level of a bihierarchy, but there is no apparent way to represent our aggregate constraints at that level.\footnote{See page 10 of \cite{akbarpour2020approximate} for the definition of the deepest level.} The random-roster construction nevertheless provides a direct route: its independent assignment of rosters across departments makes their rounding errors independent, so their university-level sum can be controlled by concentration bounds. This yields our next result.

\begin{theorem}\label{thm:approx}
There exists a randomized solution $\phi$ that is ex ante proportional, monotonic, locally proportional, and satisfies approximate aggregate proportionality.
\end{theorem}

The complete proof is in \Cref{appendix:proofthm23}. Fix a period $t$ and category $c$. Local proportionality implies that each $z^t_{d,c}$ is either $\floor{x^t_{d,c}}$ or $\ceil{x^t_{d,c}}$. Hence $A_d:=z^t_{d,c}-\floor{x^t_{d,c}}$ is a Bernoulli variable with mean $x^t_{d,c}-\floor{x^t_{d,c}}$. Because departments draw their rosters independently, the variables $A_d$ are independent, and the university-level deviation can be written as
\[
z^t_{\uni,c}-x^t_{\uni,c}=\sum_{d\in\mathcal D}\bigl(A_d-\mathbb E[A_d]\bigr).
\]
Applying Chernoff concentration bounds to this sum gives the two tail bounds in the definition of approximate aggregate proportionality. The proof also records the sharper instance-dependent rate governed by $\gamma=\sum_{d\in\mathcal D}(x^t_{d,c}-\floor{x^t_{d,c}}) <|\mathcal D|$; for deviations $b\geq |\mathcal D|$, local proportionality makes the probability exactly zero.\footnote{The property fixes the \emph{rate} at which aggregate violations become unlikely, uniformly over instances; it is not meant as a sharp probability for any given university. For a given instance, $z^t_{\uni,c}$ is a sum of $|\mathcal D|$ independent Bernoulli variables with known means, so its exact distribution (a Poisson-binomial) and hence the exact probability of any aggregate deviation can be computed directly from the cumulative vacancy counts. In the advertisement of \Cref{fig:exampleadv}, for instance, $|\mathcal D|=17$ and the largest possible aggregate deviation for any category is below 17.}

\subsection{Simpler solution for a simplified problem}\label{section:static}

\Cref{thm:impossibility} has exactly one moving part that institutions could, in principle, change: perpetual cumulative accounting (\Cref{section:monotonicity}). Suppose each recruitment cycle is treated as its own unit of account, so that compliance is assessed per cycle rather than against an ever-growing baseline. The monotonicity constraint then vanishes from the problem, along with the impossibility.

\begin{theorem} \label{thm:static}
Without the monotonicity requirement, there exists a randomized solution $\phi$ that is ex ante proportional, locally proportional, and aggregately proportional.
\end{theorem}

One way to prove this result is by utilizing the network-flow approach in decomposing the bihierarchical constraint structure of the static problem. For practical purposes, we instead prove it in \Cref{appendix: Cox} using \cite{cox1987constructive}'s controlled rounding procedure, originally devised to anonymize census tables, which is simple enough to be implemented by hand.

\Cref{thm:static} identifies a precise institutional reform under which the design conflict dissolves, and it confirms that the conflict of \Cref{thm:impossibility} is genuinely dynamic. 

It is worth noting that the running roster is not merely a bookkeeping habit. It is the device that makes reservation publicly auditable. Any citizen can open a register such as \Cref{fig:exhibit-register} and verify why a reserved point fell in one department and not another. A per-cycle lottery preserves ex ante proportionality but not this ex post traceability, and abandoning cumulative rosters would be a major departure from entrenched administrative practice. The random roster of \Cref{section:secondbest} is our recommendation \emph{within} current practice; \Cref{thm:static} shows what becomes available outside it.

\section{The design pattern beyond Indian universities}\label{section:pattern}

The ingredients of \Cref{thm:impossibility} are not specific to professors. They are (a) entitlements stated as fractions, (b) indivisible units, (c) two intersecting levels at which proportionality is expected, and (d) a running account of past allocations. Institutions with all four face exactly our problem; institutions with (a)--(c) but a different accounting convention face the static problem of \Cref{thm:static}. Four examples illustrate the range; \Cref{appendix:exhibits} reproduces the underlying documents.

\textbf{Reserve Bank of India.} The Bank recruits Class~III Assistants through a single national examination, but posts are advertised office by office. The notification reproduced in \Cref{fig:exhibit-rbi} lists 926 posts across seventeen regional offices as an office-by-category table with Bank-wide totals; bracketed entries record backlog vacancies carried forward from earlier cycles---the running account in visible form.\footnote{Because Assistants are recruited for a particular office, the SC and ST percentages applied to an office follow those prescribed for the state or region it serves rather than the national 15 and 7.5 percent; Guwahati, for example, reserves 12 of 55 posts for ST. The random-roster construction accommodates row-specific fractions: nothing in the decomposition of \Cref{appendix:proofthm2} requires the fractions to be the same across departments.} Each office maintains its own reservation register of the same form as \Cref{fig:exhibit-register}. Offices play the role of departments, the Bank that of the university, and the register is our monotonicity constraint. \Cref{thm:impossibility} applies verbatim.

\textbf{All India Services.} Officers of the Indian Administrative, Police, and Forest Services are allocated to state \emph{cadres}. For each service and each year, the Department of Personnel and Training publishes a category-wise and cadre-wise distribution of vacancies (\Cref{fig:exhibit-ias,fig:exhibit-ips,fig:exhibit-ifs}): rows are the twenty-six cadres; columns are the reservation categories, each split further into ``insider'' and ``outsider'' vacancies in a ratio fixed at 1:2.\footnote{Department of Personnel and Training, \emph{Cadre Allocation Policy for the All India Services---IAS/IPS/IFoS}, 2017, para.~1, which instructs that vacancies ``shall be divided into Unreserved/SC/ST/OBC/Insider/Outsider vacancies for each of the cadres as per the established procedure.''} The column totals track the national percentages because recruitment is national: 27 of 180 IAS vacancies for SC and 48 for OBC. Each cadre's row, however, is generated from that cadre's own running roster. Not every column total lands within a seat of its entitlement: the IPS table reserves 9 of 150 posts for ST against a fractional entitlement of 11.25, the kind of aggregate discrepancy that cadre-level rosters generate (\Cref{remark:localside}). 
This is our structure with a third dimension added, and cadre-wise counts are perpetually cumulative.

\textbf{Sri Lankan university admissions.} Places in most courses of study are filled 40 percent on island-wide merit, 55 percent through district quotas proportional to district population, and 5 percent through a quota for educationally underprivileged districts.\footnote{University Grants Commission of Sri Lanka, \emph{Admission to Undergraduate Courses of the Universities in Sri Lanka} (annual handbook). The district quota was challenged unsuccessfully in \emph{Seneviratne v.\ University Grants Commission}, [1978--79--80] 1 Sri L.R.\ 182.} Each course's district-quota places are apportioned across twenty-five districts by population share, so the admissions authority rounds a course-by-district table whose column shares are prescribed fractions---a biproportional apportionment problem with intersecting constraints. Because admissions are administered afresh each year, with unfilled places pooled and redistributed, there is no cumulative constraint. This is the static problem, and \Cref{thm:static} says that both levels can be rounded exactly and proportionally.

\textbf{Singapore's Ethnic Integration Policy.} Public housing flats are subject to ethnic-share limits at two nested levels: since 2010, at most 87 percent Chinese, 25 percent Malay, and 15 percent Indian/Other residents in a block, and 84, 22, and 12 percent in a neighbourhood \citep{wong2013estimating}. Flats are indivisible and change hands one transaction at a time, and every sale is checked against both limits. Here, however, the institution has adopted the accounting convention of case~(iii) in \Cref{section:monotonicity}: when a resident sells, the block's count for her group falls, and the limit is a cap. The dynamic problem therefore never accumulates the way a running roster does.

The four cases fall on either side of a single line. Where entitlements are cumulative and never reset---Indian universities, Reserve Bank offices, All India Service cadres---exact compliance at both levels is unattainable, and the random roster is the principled second best. Where the account is reset each cycle or adjusts with turnover---Sri Lankan admissions, Singaporean housing---exact biproportional rounding is available. The line is drawn by the accounting convention, not by the fractions or the indivisibility.

\section{Conclusion and future work}
India's decade-long struggle to reserve teaching positions simultaneously at the department and university levels was not a failure of administrative imagination. When the law prescribes fractions, positions are indivisible, allocations accrue over time, and reservations are never reversed, local proportionality, aggregate proportionality, and temporal consistency are mutually incompatible (\Cref{thm:impossibility}). The impossibility admits a principled second best: randomize the roster, comply exactly where courts require it, and concentrate tightly around the aggregate target (\Cref{thm:randomroster,thm:approx}). It has one institutional escape hatch: relax perpetual cumulative accounting, and the problem becomes easy (\Cref{thm:static}). The same line separates the other institutions in \Cref{section:pattern} into those for which the conflict is live and those for which it is not.

Several directions merit further investigation. Formal treatment of the intermediate monotonicity notions in \Cref{section:monotonicity}, including attrition-based and rolling-window accounting, would map the boundary of the impossibility more finely, as would a stochastic-arrivals analysis of the question raised in \Cref{section:scope}. An iterative rounding approach may yield solutions allowing only small additive violations of the constraints shown incompatible in \Cref{thm:impossibility} (\cite{lau2011iterative}, \cite{nguyen2019stable}), and error-minimization formulations from biproportional apportionment could be adapted to the monotonic setting (\cite{ricca2012error}). Ex post fairness and efficiency guarantees from the fair division literature may suggest richer constraint structures that coexist with monotonicity (\cite{aziz2023best}). Lastly, the multi-period considerations introduced here could be worth exploring in the classic biproportional apportionment context of translating electoral votes into parliamentary seats (\cite{balinski1989axiomatic}).

\newpage
\bibliographystyle{aer}
\bibliography{ref.bib}

@article{abdulkadirouglu2003school,
  title={School choice: A mechanism design approach},
  author={Abdulkadiro{\u{g}}lu, Atila and S{\"o}nmez, Tayfun},
  journal={American Economic Review},
  volume={93},
  number={3},
  pages={729--747},
  year={2003}
}

@article{KhannaEvren2025,
  author  = {Khanna, Manshu and Evren, Haydar},
  title   = {Rosters and connected apportionments},
  journal = {Public Choice},
  year    = {2025},
  volume  = {202},
  number  = {1--2},
  pages   = {167--191},
  doi     = {10.1007/s11127-024-01182-0},
  url     = {https://doi.org/10.1007/s11127-024-01182-0}
}

@article{nguyen2018near,
  title={Near-feasible stable matchings with couples},
  author={Nguyen, Thanh and Vohra, Rakesh},
  journal={American Economic Review},
  volume={108},
  number={11},
  pages={3154--3169},
  year={2018},
  publisher={American Economic Association 2014 Broadway, Suite 305, Nashville, TN 37203}
}

@book{sonmez2025minimalist,
  author = {S\"{o}nmez, Tayfun},
  title = {Minimalist Market Design: A Framework for Economists with Policy Aspirations},
  year = {2025},
  note = {Econometric Society Monographs Series, forthcoming},
  url = {https://arxiv.org/abs/2401.00307}
}

@article{wong2013estimating,
  author  = {Wong, Maisy},
  title   = {Estimating Ethnic Preferences Using Ethnic Housing Quotas in Singapore},
  journal = {Review of Economic Studies},
  year    = {2013}, volume = {80}, number = {3}, pages = {1178--1214}
}

@article{budish2011combinatorial,
  author  = {Budish, Eric},
  title   = {The Combinatorial Assignment Problem: Approximate Competitive Equilibrium from Equal Incomes},
  journal = {Journal of Political Economy},
  year    = {2011}, volume = {119}, number = {6}, pages = {1061--1103}
}

@article{grimmett2004stochastic,
  title={Stochastic apportionment},
  author={Grimmett, Geoffrey},
  journal={The American Mathematical Monthly},
  volume={111},
  number={4},
  pages={299--307},
  year={2004},
  publisher={Taylor \& Francis}
}

@article{nguyen2019stable,
  title={Stable matching with proportionality constraints},
  author={Nguyen, Th{\`a}nh and Vohra, Rakesh},
  journal={Operations Research},
  volume={67},
  number={6},
  pages={1503--1519},
  year={2019},
  publisher={INFORMS}
}

@book{lau2011iterative,
  title={Iterative methods in combinatorial optimization},
  author={Lau, Lap Chi and Ravi, Ramamoorthi and Singh, Mohit},
  volume={46},
  year={2011},
  publisher={Cambridge University Press}
}

@article{ehlers2014school,
  title={School choice with controlled choice constraints: Hard bounds versus soft bounds},
  author={Ehlers, Lars and Hafalir, Isa E and Yenmez, M Bumin and Yildirim, Muhammed A},
  journal={Journal of Economic theory},
  volume={153},
  pages={648--683},
  year={2014},
  publisher={Elsevier}
}

@article{bogomolnaia2001new,
  title={A new solution to the random assignment problem},
  author={Bogomolnaia, Anna and Moulin, Herv{\'e}},
  journal={Journal of Economic theory},
  volume={100},
  number={2},
  pages={295--328},
  year={2001},
  publisher={Elsevier}
}

@article{hylland1979efficient,
  title={The efficient allocation of individuals to positions},
  author={Hylland, Aanund and Zeckhauser, Richard},
  journal={Journal of Political economy},
  volume={87},
  number={2},
  pages={293--314},
  year={1979},
  publisher={The University of Chicago Press}
}

@article{aziz2023best,
  title={Best of both worlds: Ex ante and ex post fairness in resource allocation},
  author={Aziz, Haris and Freeman, Rupert and Shah, Nisarg and Vaish, Rohit},
  journal={Operations Research},
  year={2023},
  publisher={INFORMS}
}

@article{golz2022apportionment,
  title={In this apportionment lottery, the house always wins},
  author={G{\"o}lz, Paul and Peters, Dominik and Procaccia, Ariel D},
  journal={Operations Research},
  volume={74},
  number={1},
  pages={390--407},
  year={2026},
  publisher={INFORMS}
}

@article{echenique2015control,
  title={How to control controlled school choice},
  author={Echenique, Federico and Yenmez, M Bumin},
  journal={American Economic Review},
  volume={105},
  number={8},
  pages={2679--94},
  year={2015}
}

@article{ricca2012error,
  title={Error minimization methods in biproportional apportionment},
  author={Ricca, Federica and Scozzari, Andrea and Serafini, Paolo and Simeone, Bruno},
  journal={Top},
  volume={20},
  number={3},
  pages={547--577},
  year={2012},
  publisher={Springer}
}

@article{budish2013designing,
  title={Designing random allocation mechanisms: Theory and applications},
  author={Budish, Eric and Che, Yeon-Koo and Kojima, Fuhito and Milgrom, Paul},
  journal={American Economic Review},
  volume={103},
  number={2},
  pages={585--623},
  year={2013}
}

@book{balinski2010fair,
  title={Fair representation: meeting the ideal of one man, one vote},
  author={Balinski, Michel L and Young, H Peyton},
  year={2010},
  publisher={Brookings Institution Press}
}

@article{dur2019explicit,
  title={Explicit vs. Statistical Preferential Treatment in Affirmative Action: Theory and Evidence from Chicago’s Exam Schools.” forthcoming},
  author={Dur, Umut and Pathak, Parag and S{\"o}nmez, Tayfun},
  journal={Journal of Economic Theory},
  year={2019}
}

@Article{balinski1989axiomatic,
  author    = {Balinski, Michel L and Demange, Gabrielle},
  title     = {An axiomatic approach to proportionality between matrices},
  journal   = {Mathematics of Operations Research},
  year      = {1989},
  volume    = {14},
  number    = {4},
  pages     = {700--719},
  publisher = {INFORMS},
}

@article{sonmez2019affirmative,
author = {Sönmez, Tayfun and Yenmez, M. Bumin},
title = {Affirmative Action in India via Vertical, Horizontal, and Overlapping Reservations},
journal = {Econometrica},
volume = {90},
number = {3},
pages = {1143-1176},
doi = {https://doi.org/10.3982/ECTA17788},
url = {https://onlinelibrary.wiley.com/doi/abs/10.3982/ECTA17788},
eprint = {https://onlinelibrary.wiley.com/doi/pdf/10.3982/ECTA17788},
year = {2022}
}

@book{pukelsheim2017proportional,
  title={Proportional representation},
  author={Pukelsheim, Friedrich},
  year={2017},
  publisher={Springer}
}

@Article{akbarpour2020approximate,
  author  = {Akbarpour, Mohammad and Nikzad, Afshin},
  title   = {Approximate random allocation mechanisms},
  journal = {The Review of Economic Studies},
  year    = {2020},
}

@Article{cox1987constructive,
  author    = {Cox, Lawrence H},
  title     = {A constructive procedure for unbiased controlled rounding},
  journal   = {Journal of the American Statistical Association},
  year      = {1987},
  volume    = {82},
  number    = {398},
  pages     = {520--524},
  publisher = {Taylor \& Francis Group},
}

@article{chernoff1952measure,
  title={A measure of asymptotic efficiency for tests of a hypothesis based on the sum of observations},
  author={Chernoff, Herman},
  journal={The Annals of Mathematical Statistics},
  pages={493--507},
  year={1952},
  publisher={JSTOR}
}

@incollection{edmonds2003submodular,
  title={Submodular functions, matroids, and certain polyhedra},
  author={Edmonds, Jack},
  booktitle={Combinatorial Optimization—Eureka, You Shrink!},
  pages={11--26},
  year={2003},
  publisher={Springer}
}

@Article{Aygun2021,
  author  = {Ayg{\"u}n, Orhan and B{\'o}, In{\'a}cio},
  journal = {American Economic Journal: Microeconomics},
  title   = {College admission with multidimensional privileges: The Brazilian affirmative action case},
  year    = {2021},
  number  = {3},
  pages   = {1--28},
  volume  = {13},
}

@Article{Ayguen2017,
  author  = {Ayg{\"u}n, Orhan and Turhan, Bertan},
  journal = {American Economic Review},
  title   = {Large-scale affirmative action in school choice: Admissions to IITs in India},
  year    = {2017},
  number  = {5},
  pages   = {210--13},
  volume  = {107},
}
\newpage
\crefalias{subsection}{appendix}
\appendix

\section*{Appendices for Online Publication}

\section{Proofs} \label{Appendix: Proofs}

\subsection{Proof of \Cref{thm:randomroster}}\label{appendix:proofthm2}

Since the sequence of cumulative fractional assignments is unknown, the solution must work for any possible sequence. The essential technique is to design a procedure that takes the fractions of reservations and generates a roster for a \textit{single} department. Local proportionality regulates the cumulative number of vacancies for each category in a department's roster. Since many rosters could be locally proportional, the procedure generates a \textit{random roster} by assigning each solution roster a probability.

\begin{proof}[Proof of \Cref{thm:randomroster}]

In this proof, we will construct a set of rosters for departments that ensure the integral assignment is locally proportional for any sequence of fractional assignments.

\noindent \textbf{Part 1: Constructing a roster for departments.}

Let $k$ be the least common multiple of the denominators of the reservation fractions in $\bm{\upalpha}$. Since the $\alpha_c$'s are rational numbers, such $k$ exists. Let $P$ represent the given reservation scheme as a $k \times |\mathcal{C}|$ two-way table, where the rows denote the index of the seats and the columns denote the categories. The internal entry $p_{ic}$ equals $\alpha_{c}$ for every $(i,c)$.\footnote{We are using new notation because this two-way table $P$ must not be confused with the fractional assignment table $X^t$; here $i$ indexes seats, not departments.}

We want to transform this table of fractional entries into an integral table. The integral table would assign each of the $k$ vacancies (in rows) to a category (in columns); that is, it would generate a unique roster.

We introduce some definitions for the constraints that monotonicity and local proportionality impose on the table $P$. An \textbf{element} $(i,c)$ denotes an entry in row $i$ and column $c$ of $P$. Let $E$ be the set of all elements in $P$. A \textbf{constraint} is a subset $K \subseteq E$ of elements with a floor $\floor{\sum_{(i,c)\in K} p_{ic}}$ and a ceiling $\ceil{\sum_{(i,c)\in K} p_{ic}}$. A \textbf{constraint set} is a set of constraints. A \textbf{constraint structure} is a set of subsets of $E$.

We are looking for integral tables $P'$ that satisfy the following three types of constraints. \Cref{Appendix: A.1} gives an example to help understand the constraint structure better.

\begin{enumerate}

\item Internal constraints ensure that each internal entry can be 1 or 0. \\That is, $0 \leq p'_{ic} \leq 1$ for every $(i,c)$. \\ Let $\mathcal{K}_I$ be the constraint structure associated with internal constraints.

\item Row sums are required to be one since every position is assigned to exactly one category.\\ That is, $ \sum_{c \in \mathcal{C}} p'_{ic} =1$ for every $i$. \\ Let $\mathcal{K}_R$ be the constraint structure associated with row constraints.

\item Column constraints ensure that the difference between the cumulative sum of vacancies given to a category and cumulative fractional assignments is less than one.\\ That is, $\floor{\sum_{i \leq l} p_{ic}} \leq \sum_{i \leq l} p'_{ic} \leq \ceil{\sum_{i \leq l} p_{ic}}$ for every $2\leq l \leq k$ and $c \in \mathcal{C}$.  \\ Let $\mathcal{K}_C$ be the constraint structure associated with column constraints.

\end{enumerate}

Therefore, we will consider tables $P'$ that satisfy, for each $K \in \mathcal{K}_C \cup \mathcal{K}_I \cup \mathcal{K}_R$, $$\floor{\sum_{(i,c)\in K} p_{ic}}  \leq \sum_{(i,c)\in K} p'_{ic} \leq \ceil{\sum_{(i,c)\in K} p_{ic}}.$$

Notice that each of the constraint structures associated with the constraint sets $\mathcal{K}_C \cup \mathcal{K}_I$ and $\mathcal{K}_R \cup \mathcal{K}_I$ is a hierarchy. Recall from \cite{budish2013designing}, a constraint structure $\mathcal{K}$ is a \textbf{hierarchy} if for every pair $E$ and $E'$ in $\mathcal{K}$, we have that $E\subset E'$ or $E' \subset E$ or $E \cap E' = \emptyset$. Therefore, due to \cite{edmonds2003submodular} and \cite{budish2013designing}, an ex ante proportional randomized solution exists that randomizes over integral assignment tables satisfying the above constraints. The technique to construct such a solution is further detailed in \Cref{Appendix: A.2} and \Cref{Appendix: A.2.2}.

In particular, by Theorem 1 of \cite{budish2013designing}, there exist positive numbers $\lambda_1, \ldots, \lambda_N$, summing to one, and integral assignments $\bar{P}_1, \ldots, \bar{P}_N$, feasible under the constraints in $\mathcal{K}_C \cup \mathcal{K}_I \cup \mathcal{K}_R$, such that
$$ P = \sum_{h=1}^{N} \lambda_h \bar{P}_h .$$

{Generating rosters:} Recall that a roster pins down an allocation of an infinite sequence of seats constructed as a concatenation of infinitely many finite seat sequences of length $k$. Generate a unique roster $\sigma_h$ for each $\bar{P}_h$ such that $\bar{p}_{ic} =1$ implies $\sigma_h(\ell k+i)=c$ for every non-negative integer $\ell$.

\noindent \textbf{Part 2: Assigning rosters to departments independently.}

Independently assigning each department a roster $\sigma_h$ with probability $\lambda_h$ generates the required randomized solution $\phi^*$. Each department then reserves vacancies in every period according to the assigned roster: if roster $\sigma$ is realized for department $d$, the vacancies reserved in $d$ in period-1 are determined by $\sigma(1),\dotso, \sigma(q^1_d)$; in period-2 by $\sigma(q^1_d +1),\dotso, \sigma(q^1_d + q^2_d)$; and so forth.

By construction, $\phi^*$ is locally proportional: ex post, department-level fractional assignments are rounded up or down. The solution is monotonic because each department follows a roster that can only weakly increase each category's cumulative integer assignment; these assignments sum up at the university level, making the solution monotonic there too. Moreover, $\phi^*$ is ex ante proportional: the expected number of vacancies reserved for category $c$ in department $d$ until period-$t$ is $\mathbb{E}(z_{d,c}^t)=\sum_{s\leq t}q^{s}_{d} \alpha_{c}$, which equals the internal entry $x^t_{d,c}$ of $X^t$. Thus $\mathbb{E}_\phi[Z^t] = X^t$. This proves the theorem.
\end{proof}

\begin{proof}[Proof of \Cref{remark:universityversion}]
Replacing Part 2 by assigning roster $\sigma_h$ with probability $\lambda_h$ to the university (instead of to departments) generates a randomized solution $\phi'$ that is ex ante proportional, monotonic, and aggregately proportional.
\end{proof}

\subsection{Proof of \Cref{thm:approx}}\label{appendix:proofthm23}

\begin{proof}[Proof of \Cref{thm:approx}] We will prove that the randomized solution $\phi^*$ of \Cref{thm:randomroster} satisfies approximate aggregate proportionality. We show this by proving two lemmas: first, that entries of each column of $Z^t$ are independent; second, that the Chernoff concentration bounds then apply.

\begin{lemma}\label{lemma 1}
For any subset $S \subset \mathcal{D}$ and any $c \in \mathcal{C}$, we have
$$ \text{Pr}\Big[ \bigwedge_{d \in S} z^t_{d,c} = \ceil{x^t_{d,c}} \Big] = \prod_{d \in S} \text{Pr}\Big[ z^t_{d,c}= \ceil{x^t_{d,c}} \Big] \text{ ,} $$ 
$$\text{Pr}\Big[ \bigwedge_{d \in S} z^t_{d,c} = \floor{x^t_{d,c}} \Big] = \prod_{d \in S} \text{Pr}\Big[ z^t_{d,c}= \floor{x^t_{d,c}} \Big] \text{ .}$$
\end{lemma}
\begin{proof}
The random roster is assigned to each department independently. Consequently, the variables $\{z^t_{d,c}\}_{d\in D}$ are mutually independent.
\end{proof}

We next recall a result of \cite{chernoff1952measure}. Chernoff bounds are well-known concentration inequalities that limit the deviation of a sum of independent Bernoulli random variables from its mean; here we use the multiplicative form.

\begin{lemma}[Chernoff bounds]\label{lemma3}
Let \(A_1,A_2,\ldots,A_{|D|}\) be independent random variables taking values
in \(\{0,1\}\), and let \(\mu=\sum_{d\in D}E[A_d]\). For every
\(\varepsilon\geq 0\),
\[
\Pr\left[\sum_{d\in D}A_d\geq (1+\varepsilon)\mu\right]
\leq
\exp\left(-\frac{\mu\varepsilon^2}{2+\varepsilon}\right).
\]
For every \(0\leq\varepsilon\leq 1\),
\[
\Pr\left[\sum_{d\in D}A_d\leq (1-\varepsilon)\mu\right]
\leq
\exp\left(-\frac{\mu\varepsilon^2}{2}\right).
\]
\end{lemma}

\begin{lemma}\label{lemma 2}
Suppose that the random variables \(z^t_{d,c}\), for \(d\in D\) and
\(c\in C\), are such that
\begin{enumerate}
    \item \(E[z^t_{d,c}]=x^t_{d,c}\), 
    \item each \(z^t_{d,c}\) takes only the two values
    \(\lfloor x^t_{d,c}\rfloor\) and \(\lceil x^t_{d,c}\rceil\), and
    \item the $z^t_{d,c}$ are independent across $d$.
\end{enumerate}
Let
\[
\gamma=\sum_{d\in D}\left(x^t_{d,c}-\lfloor x^t_{d,c}\rfloor\right).
\]
Then, for every \(b>0\),
\[
\Pr\left[\sum_{d\in D}(z^t_{d,c}-x^t_{d,c})\geq b\right]
\leq
\exp\left(-\frac{b^2}{2\gamma+b}\right),
\]
and
\[
\Pr\left[\sum_{d\in D}(z^t_{d,c}-x^t_{d,c})\leq -b\right]
\leq
\begin{cases}
\exp\left(-\dfrac{b^2}{2\gamma}\right),&0<b\leq\gamma,\\[6pt]
0,&b>\gamma.
\end{cases}
\]
\end{lemma}

\begin{proof}
Fix \(c\in C\), and define
\[
A_d:=z^t_{d,c}-\lfloor x^t_{d,c}\rfloor.
\]
By local proportionality, \(A_d\in\{0,1\}\). By Lemma 1, the variables
\(A_d\) are independent. Moreover,
\[
E[A_d]=x^t_{d,c}-\lfloor x^t_{d,c}\rfloor,
\qquad
\gamma=\sum_{d\in D}E[A_d],
\]
and
\[
z^t_{d,c}-x^t_{d,c}=A_d-E[A_d].
\]

If \(\gamma=0\), then \(A_d=0\) almost surely for every \(d\), so the
aggregate deviation is zero almost surely and the bounds hold. Suppose
therefore that \(\gamma>0\).

For the upper tail, let \(\varepsilon=b/\gamma\). The upper Chernoff bound
from Lemma 2 gives
\[
\begin{aligned}
\Pr\left[\sum_{d\in D}(z^t_{d,c}-x^t_{d,c})\geq b\right]
&=
\Pr\left[\sum_{d\in D}A_d\geq \gamma+b\right]\\
&=
\Pr\left[\sum_{d\in D}A_d\geq (1+\varepsilon)\gamma\right]\\
&\leq
\exp\left(-\frac{\gamma\varepsilon^2}{2+\varepsilon}\right)\\
&=
\exp\left(-\frac{b^2}{2\gamma+b}\right).
\end{aligned}
\]

For the lower tail, if \(b>\gamma\), then
\[
\sum_{d\in D}(A_d-E[A_d])\geq-\gamma>-b
\]
almost surely, so the probability is zero. If \(0<b\leq\gamma\), let
\(\varepsilon=b/\gamma\in(0,1]\). The lower Chernoff bound gives
\[
\begin{aligned}
\Pr\left[\sum_{d\in D}(z^t_{d,c}-x^t_{d,c})\leq-b\right]
&=
\Pr\left[\sum_{d\in D}A_d\leq\gamma-b\right]\\
&=
\Pr\left[\sum_{d\in D}A_d\leq(1-\varepsilon)\gamma\right]\\
&\leq
\exp\left(-\frac{\gamma\varepsilon^2}{2}\right)
=
\exp\left(-\frac{b^2}{2\gamma}\right).
\end{aligned}
\]
\end{proof}

For a locally proportional solution, the deviation at the university level is bounded above, as $\left|z_{\uni,c}^t-x_{\uni,c}^t\right|  \leq \sum_{d} \left|z^t_{d,c} - x^t_{d,c}\right|\leq \sum_{d} \max\{x^t_{d,c}-\floor{x^t_{d,c}},\ceil{x^t_{d,c}}-x^t_{d,c}\}  < |\mathcal{D}|$. Therefore, for $b\geq |\mathcal{D}|$,
$$ \text{Pr}\Big[ \left| z_{\uni,c}^t-x_{\uni,c}^t \right| \geq  b \Big] =0 \text{ .}$$

Since every fractional part is strictly less than one,
\(\gamma<|D|\). Hence, for \(0<b<|D|\),
\[
2\gamma+b<3|D|,
\]
and therefore
\[
\exp\left(-\frac{b^2}{2\gamma+b}\right)
<
\exp\left(-\frac{b^2}{3|D|}\right).
\]
For \(0<b\leq\gamma\),
\[
\exp\left(-\frac{b^2}{2\gamma}\right)
<
\exp\left(-\frac{b^2}{2|D|}\right),
\]
while for \(b>\gamma\) the lower-tail probability is zero.

Therefore \(\phi^*\) satisfies approximate aggregate proportionality.
\end{proof}

\subsection{An example for \Cref{thm:randomroster}: Explaining constraints} \label{Appendix: A.1}

To understand the constraint structure of the problem better, take the following example. Consider a problem with two categories $\mathcal{C}=\{c_1, c_2 \}$ and the reservation scheme $\bm{\upalpha} = (\alpha_{1},\alpha_{2})=(1/3,2/3)$. We represent the given reservation scheme as a two-way table $P$, where the rows denote the index of the vacancies and the columns denote the categories. Each internal entry $p_{ic}=\alpha_c$. Take $k=3$, the least common multiple of the denominators.

\Cref{fig:thm1} illustrates the constraint structure. Column constraints are $C_{31}=\{k_{11},k_{21},k_{31}\}$, $C_{21}=\{k_{11},k_{21}\}$, $C_{32}=\{k_{12},k_{22},k_{32}\}$, and $C_{22}=\{k_{12},k_{22}\}$, and row constraints are $R_1=\{k_{11},k_{12}\}$, $R_2=\{k_{21},k_{22}\}$, and $R_3=\{k_{31},k_{32}\}$.

\newcounter{nodecount}
\newcommand\tabnode[1]{\addtocounter{nodecount}{1} \tikz \node  (\arabic{nodecount}) {#1};}
\tikzstyle{every picture}+=[remember picture,baseline]
\tikzstyle{every node}+=[anchor=base,
minimum width=1.8cm,align=center,text depth=.25ex,outer sep=1.5pt]
\tikzstyle{every path}+=[thick, rounded corners]

\begin{figure}[htb]
\vspace{0.6in}
\centering

\begin{subfigure}{.4\linewidth}
\textbf{$P$} =
\begin{tabular}{cc|c}
\tabnode{1/3} & \tabnode{2/3} & 1\tabularnewline
\tabnode{1/3} & \tabnode{2/3} & 1\tabularnewline
\tabnode{1/3} & \tabnode{2/3} & 1\tabularnewline
\hline 
1 & 2 &3\tabularnewline
\end{tabular}

\begin{tikzpicture}[overlay]

\node[draw=black,rounded corners = 1ex,fit=(1)(1),inner sep = 0pt] {};
\node [right=0cm,above=1cm,minimum width=0pt] at (1.west) (14) {{$k_{11}$}};
\draw [<-,out=90,in=270, black] (1.north) to (14);

\node[draw=black,rounded corners = 1ex,fit=(3)(3),inner sep = 0pt] {};
\node [right=0cm,above=1.1cm,minimum width=0pt] at (1.east) (15) {{$k_{21}$}};
\draw [<-,out=90,in=270, black] (3.north) to (15);

\node[draw=black,rounded corners = 1ex,fit=(5)(5),inner sep = 0pt] {};

\node [right=0cm,below=1cm,minimum width=0pt] at (5.west) (16) {{$k_{31}$}};
\draw [<-,out=270,in=90, black] (5.south west) to (16);

\node[draw=black,rounded corners = 1ex,fit=(2)(2),inner sep = 0pt] {};
\node [right=0cm,above=0.7cm,minimum width=0pt] at (2.west) (17) {{$k_{12}$}};
\draw [<-,out=90,in=270, black] (2.north) to (17);

\node[draw=black,rounded corners = 1ex,fit=(4)(4),inner sep = 0pt] {};
\node [right=0cm,above=1cm,minimum width=0pt] at (2.east) (18) {{$k_{22}$}};
\draw [<-,out=90,in=270, black] (4.north) to (18);

\node[draw=black,rounded corners = 1ex,fit=(6)(6),inner sep = 0pt] {};
\node [right=0cm,below=1cm,minimum width=0pt] at (6.east) (19) {{$k_{32}$}};
\draw [<-,out=270,in=90, black] (6.south east) to (19);

\end{tikzpicture}
\vspace{0.3in}
\caption{INTERNAL CONSTRAINTS}
\end{subfigure}\\[0.7in]

\centering
  \begin{subfigure}{.4\linewidth}
        \setcounter{nodecount}{0}
        \centering
        \textbf{$P$} =
        \begin{tabular}{cc|c}
        \tabnode{1/3} & \tabnode{2/3} & 1\tabularnewline
        \tabnode{1/3} & \tabnode{2/3} & 1\tabularnewline
        \tabnode{1/3} & \tabnode{2/3} & 1\tabularnewline
        \hline 
        1 & 2 &3\tabularnewline
        \end{tabular}
    
        \begin{tikzpicture}[overlay]
        
        \node[draw=black,rounded corners = 1ex,fit=(1)(5),inner sep = 2pt] {};
        \node [right=0cm,below=1cm,minimum width=0pt] at (5.west) (7) {\textcolor{black}{$C_{31}$}};
        \draw [<-,out=270,in=90, black] (5.south west) to (7);
        
        \node[draw=black,rounded corners = 1ex,fit=(1)(3),inner sep = 0pt] {};
        \node [right=2cm,above=1.5cm,minimum width=0pt] at (3) (8) {\textcolor{black}{$C_{21}$}};
        \draw [<-,out=90,in=270, black] (1) to (8);
        
        \node[draw=black,rounded corners = 1ex,fit=(2)(6),inner sep = 2pt] {};
        \node [right=2cm,below=1cm,minimum width=0pt] at (6.east) (9) {\textcolor{black}{$C_{32}$}};
        \draw [<-,out=270,in=90, black] (6.south east) to (9);
        
        \node[draw=black,rounded corners = 1ex,fit=(2)(4),inner sep = 0pt] {};
        \node [right=2cm,above=1.5cm,minimum width=0pt] at (4) (10) {\textcolor{black}{$C_{22}$}};
        \draw [<-,out=90,in=270, black] (2) to (10);

\end{tikzpicture}
\vspace{0.3in}
\caption{COLUMN CONSTRAINTS}    
    
  \end{subfigure}%
  \hspace{1em}%
  \begin{subfigure}{.4\linewidth}
  \setcounter{nodecount}{0}
        \centering
        \textbf{$P$} =
         \begin{tabular}{cc|c}
        \tabnode{1/3} & \tabnode{2/3} & 1\tabularnewline
        \tabnode{1/3} & \tabnode{2/3} & 1\tabularnewline
        \tabnode{1/3} & \tabnode{2/3} & 1\tabularnewline
        \hline 
        1 & 2 &3\tabularnewline
        \end{tabular}
    
\begin{tikzpicture}[overlay]

\node[draw=black,rounded corners = 1ex,fit=(1)(2),inner sep = 0pt] {};
\node [right=0cm,above=1cm,minimum width=0pt] at (2) (11) {\textcolor{black}{$R_1$}};
\draw [<-,out=90,in=270, black] (2) to (11);

\node[draw=black,rounded corners = 1ex,fit=(3)(4),inner sep = 0pt] {};
\node [right=0cm,above=1cm,minimum width=0pt] at (1) (12) {\textcolor{black}{$R_2$}};
\draw [<-,out=90,in=270, black] (3) to (12);

\node[draw=black,rounded corners = 1ex,fit=(5)(6),inner sep = 0pt] {};
\node [right=0cm,below=1cm,minimum width=0pt] at (5.west) (13) {\textcolor{black}{$R_3$}};
\draw [<-,out=270,in=90, black] (5.south west) to (13);

\end{tikzpicture}
\vspace{0.3in}
\caption{ROW CONSTRAINTS}   
\end{subfigure}%

\caption{CONSTRAINT STRUCTURE OF THE EXAMPLE $P$}
\label{fig:thm1}

\end{figure}

\subsection{Decomposition to construct rosters in \Cref{thm:randomroster}} \label{Appendix: A.2}

\noindent For practical purposes, we provide a method that generates rosters in polynomial time.

\noindent \textbf{Part 1. Revisiting the constraints.}\\
Recall that in the proof of \Cref{thm:randomroster} we were looking for integral tables $P'$ that satisfy three types of constraints:

\begin{enumerate}

\item Internal constraints ensure that each internal entry can be 1 or 0. \\That is, $0 \leq p'_{ic} \leq 1$ for every $(i,c)$. \\ Let $\mathcal{K}_I$ be the constraint structure associated with internal constraints. \\Let $k_{ic}:=\{ (i,c)\}$ denote such a constraint.

\item Row sums are required to be one since every position is assigned to exactly one category.\\ That is, $ \sum_{c \in \mathcal{C}} p'_{ic} =1$ for every $i$. \\ Let $\mathcal{K}_R$ be the constraint structure associated with row constraints. \\Let $R_i:=\{ (i,c) \mid c \in \mathcal{C} \}$ denote such a constraint.

\item Column constraints ensure that the difference between the cumulative sum of vacancies given to a category and cumulative fractional assignments is less than one.\\ That is, $\floor{\sum_{i \leq l} p_{ic}} \leq \sum_{i \leq l} p'_{ic} \leq \ceil{\sum_{i \leq l} p_{ic}}$ for every $2\leq l \leq k$ and $c \in \mathcal{C}$.  \\ Let $\mathcal{K}_C$ be the constraint structure associated with column constraints. \\Let $C_{lc}:= \{ (i,c) \mid i \leq l \}$ denote such a constraint.

\end{enumerate}

\noindent \textbf{Part 2. Map table $P$ to a flow network.}\\
We will now map table $P$ with the above constraint structure to a flow network. The set of vertices consists of the source, the sink, vertices for each $k_{ic} \in \mathcal{K}_I$, each $R_i \in \mathcal{K}_R$, and each $C_{lc} \in \mathcal{K}_C$. The following rules govern the placement of directed edges:

\begin{enumerate}
    \item A directed edge from source to $C_{kc}$ for every $c \in \mathcal{C}$.
    \item A directed edge from $C_{lc}$ to $k_{lc}$ and $C_{l-1,c}$ for every $l \geq 3$ and $c \in \mathcal{C}$. 
    \item A directed edge from $C_{2c}$ to $k_{2c}$ and $k_{1c}$ for every $c \in \mathcal{C}$.
    \item A directed edge from $k_{ic}$ to $R_i$ for every $(i,c)$.
    \item A directed edge from $R_i$ to sink for every $i$.
\end{enumerate}

We next associate flow with each edge. Notice that there is only one incoming edge for each vertex $K \in \mathcal{K}_C \cup \mathcal{K}_I$, and only one outgoing edge for each vertex $K \in \mathcal{K}_R \cup \mathcal{K}_I$, because of the hierarchical constraint structure. For each vertex $K \in \mathcal{K}_C \cup \mathcal{K}_I$, the incoming flow equals $\sum_{(i,c)\in K} p_{ic}$; for each vertex $K \in \mathcal{K}_R \cup \mathcal{K}_I$, the outgoing flow equals $\sum_{(i,c)\in K} p_{ic}$. The flow association ensures that incoming flow equals outgoing flow at each vertex.

We have mapped table $P$ with the constraint structures to a flow network. The mapping is injective; as long as the constraints are still satisfied after the transformation, every transformation in the flow network can be mapped back to table $P$.

\medskip

\noindent \textbf{Part 3. Generate integral assignments for rosters.}\\
We call the pair of tables $(P^1, P^2)$ a \textbf{decomposition} of table $P$ if
\begin{enumerate}
    \item there exists $\beta \in (0,1)$ such that $P=\beta P^1 + (1-\beta) P^2$,
    \item for each constraint $K$, $\floor{\sum_{(i,c)\in K} p_{ic}}  \leq \sum_{(i,c)\in K} p^l_{ic} \leq \ceil{\sum_{(i,c)\in K} p_{ic}}$ for $l=1,2$, and
    \item tables $P^1$ and $P^2$ have more integral entries than table $P$.
\end{enumerate}
\vspace{0.1in}
\textbf{Decomposition Algorithm}
\begin{head}
If the flow network contains a fractional edge:\\
\textbf{Step 1:} Choose any edge that has fractional flow. Since the total inflow equals the total outflow for each vertex, there will be an adjacent edge with fractional flow.\footnote{The reason is that an integer cannot be written as the sum of integers and a proper fraction.} Continue to add new edges with fractional flows until a cycle is formed.\footnote{The number of vertices is finite, so we can find a cycle.}\\
\textbf{Step 2:} Modify the flows in the cycle in two ways to create $P^1$ and $P^2$:
\begin{enumerate}[nosep,topsep=-1ex,leftmargin=9\labelsep,rightmargin=4\labelsep]
        \item First way: the flow of each forward edge\footnote{We say an edge in the cycle is a forward edge if its direction is the same as the cycle path, the backward edge is defined conversely.} is increased and the flow of each backward edge is decreased at the same rate until at least one flow reaches an integer value. Record the amount of adjustment as $d_{+}$. Map back the resulting flow network to a two-way table. Denote the table as $P^1$.
        \item Second way: the flow of each forward edge is decreased, and the flow of each backward edge is increased at the same rate until at least one flow reaches an integer value. Record the amount of adjustment as $d_{-}$. Map back the resulting flow network to a two-way table. Denote the table as $P^2$.
        \item Set $\beta =\frac{d_{-}}{d_{-}+d_{+}}$.
        \item The pair of tables $(P^1, P^2)$ is a decomposition of table $P$, where $P=\beta P^1 + (1-\beta) P^2$.
    \end{enumerate}

\end{head}\vspace{0.1in}

Generate the integral assignment tables $\bar{P}_1, \ldots, \bar{P}_N$ with probabilities $\lambda_1, \ldots, \lambda_N$ in the following way. Decompose $P$ using the decomposition algorithm into a convex combination $P=\beta P^1 + (1-\beta) P^2$ of two assignment tables, each of which has at least one more integer-valued element. Then generate a random number and, with probability $\beta$, continue by similarly decomposing $P^1$, while with probability $1-\beta$, continue by decomposing $P^2$. Stop at an integral assignment.

This generates the integral two-way tables with the same constraint structure as table $P$. Moreover, the (compound) probability of arriving at each integral table is such that the expected assignment table equals $P$.

\subsection{Example for \Cref{thm:randomroster} continued: Explaining decomposition} \label{Appendix: A.2.2}
\noindent The example's table $P$ is 

\begin{figure}[!htb]
  \centering
        \textbf{$P$} =
        \begin{tabular}{cc|c}
        1/3 & 2/3 & 1\tabularnewline
        1/3 & 2/3 & 1\tabularnewline
        1/3 & 2/3 & 1\tabularnewline
        \hline 
        1 & 2 & 3\tabularnewline
    \end{tabular}
\end{figure}

\begin{figure}[!htb]
    \centering
    \includegraphics[width=1\textwidth]{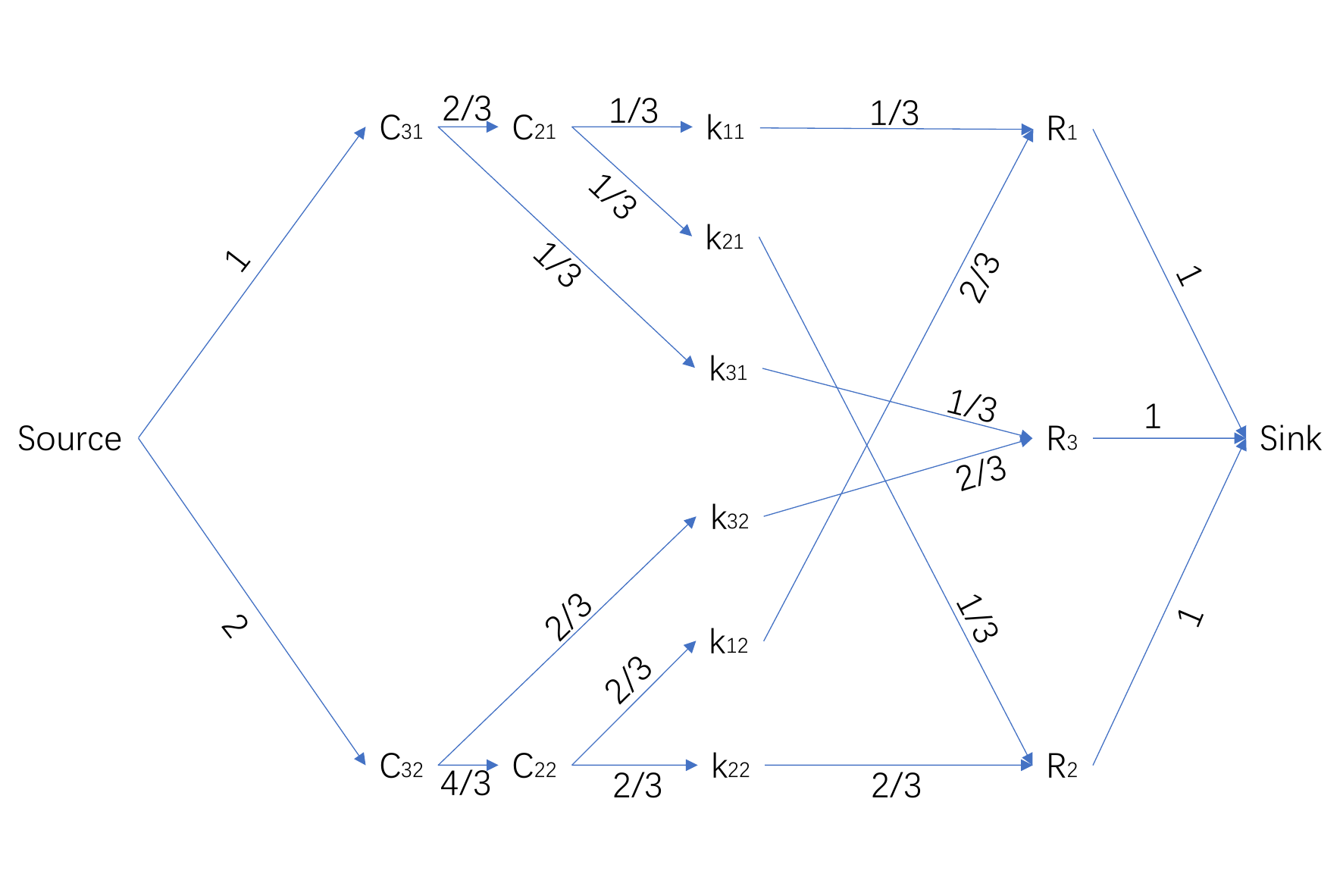}
    
    \caption{FLOW NETWORK REPRESENTATION OF THE EXAMPLE $P$}
    \label{fig:fig8}
\end{figure}

The two-way table $P$ with the constraints is then represented as a network flow. Starting from the source, the flows pass through the sets in column constraints, arranged in descending order of set inclusion. That is, for example, $C_{31} \supset C_{21} \supset k_{11}$. This explains the flow network on the left side of \Cref{fig:fig8}, where the numbers on the edges represent the flows. The flows then proceed along the directed edges representing the set-inclusion tree, eventually reaching the singleton sets. That is, for example, $k_{11} \subset R_1$. This explains the flow network on the right side of \Cref{fig:fig8}.

In the flow network, note that the flow associated with each edge reflects the totals of elements in the corresponding set, and the flow arriving at each vertex equals the flow leaving that vertex. The algorithm will conserve these two properties while constructing a new flow network with fewer fractional elements.

\begin{figure}[h]
    \centering
    \includegraphics[width=1\textwidth]{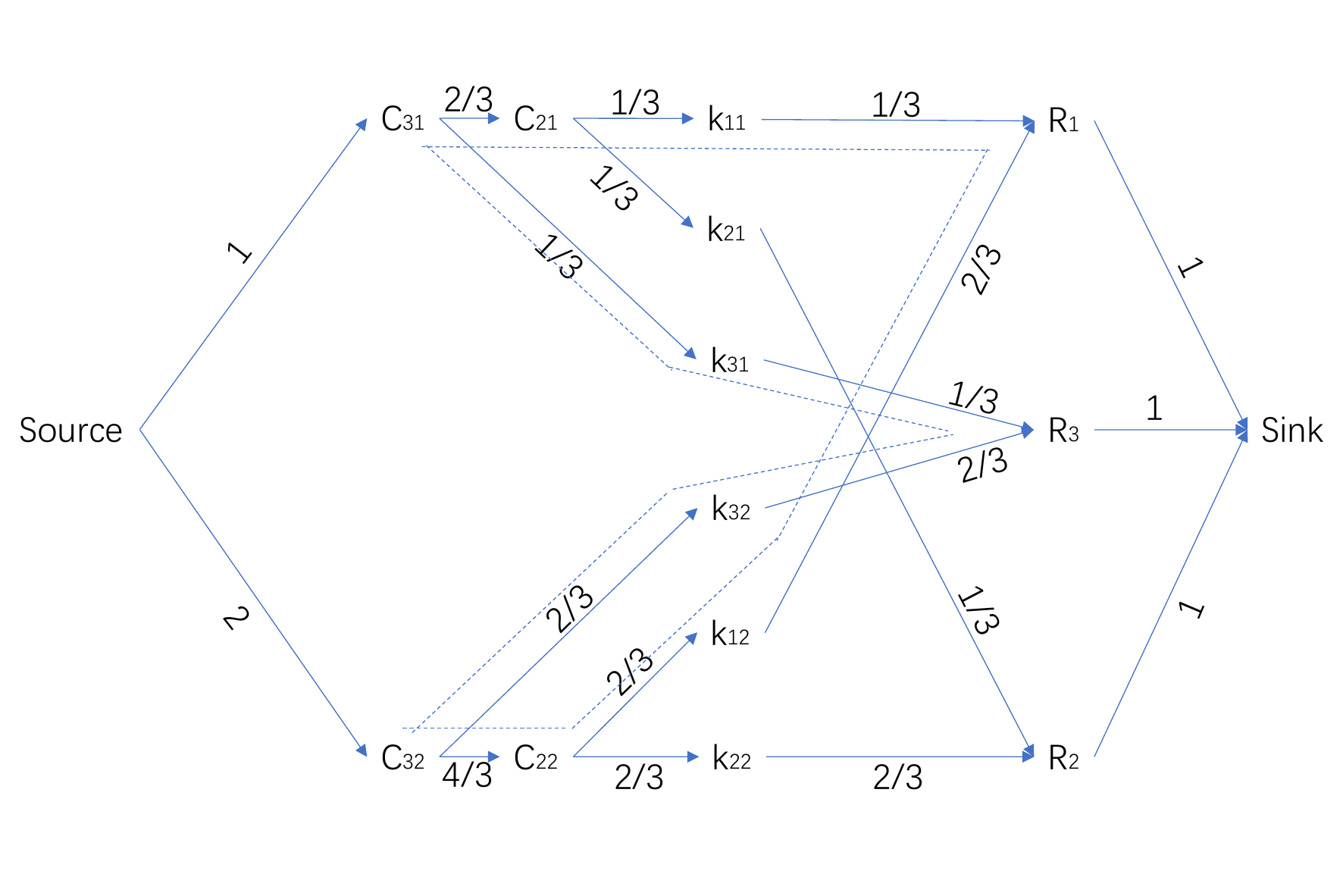}
    
    \caption{AN EXAMPLE OF CYCLE WITH FRACTIONAL EDGES}
    \label{fig:fig9}
\end{figure}

We first identify a cycle of edges with fractional flows. Choosing any fractional edge, say $(C_{31}, k_{31})$, we find another fractional edge that is a neighbor of $k_{31}$. If a vertex has a fractional edge, it has to have another fractional edge: since total inflow equals outflow for every vertex (except source and sink), we would otherwise have a contradiction. We continue to add new fractional edges until we form a cycle. In our example, the cycle of fractional edges is $C_{31}\rightarrow^{1/3} k_{31} \rightarrow^{1/3} R_3 \leftarrow^{2/3}  k_{32} \leftarrow^{2/3} C_{32} \rightarrow^{4/3} C_{22} \dotso  \leftarrow^{2/3} C_{31}$. We illustrate this cycle in \Cref{fig:fig9} with dashed lines.

Next, we alter the cycle's edge flows. We first increase each forward edge's flow while simultaneously decreasing each backward edge's flow until at least one flow reaches an integer value. A table $P_1$ is created from the resulting network flow. In the example, flows along all forward edges increase from 2/3 to 1, 1/3 to 2/3, and 4/3 to 5/3, while flows along all backward edges decrease from 1/3 to 0 and 2/3 to 1/3. The adjustment is $d_{+}=1/3$. Next, the flows of the edges in the cycle are readjusted in the opposite direction, increasing those with backward edges and lowering those with forward edges analogously, resulting in a new table $P_2$. In the example, flows along all forward edges decrease from 2/3 to 1/3, 1/3 to 0, and 4/3 to 1, while flows along all backward edges increase from 1/3 to 2/3 and 2/3 to 1. The adjustment is $d_{-}=1/3$.

Now, we can decompose $P$ into these two tables: $P=\frac{d_{-}}{d_{-}+d_{+}} P_1 + \frac{d_{+}}{d_{-}+d_{+}} P_2=\frac{1}{2} P_1 + \frac{1}{2} P_2$. The algorithm picks $P_1$ with probability 0.5 and $P_2$ with probability 0.5. We reiterate the decomposition process until no fractions are left.

At least one fraction in $P$ is converted to an integer at each iteration, while all current integers remain constant. Each fraction must appear in at least one iteration. As a result, the process must converge to an integer table in fewer iterations than the initial number of fractions in table $P$.

Since only the fractions along one cycle in the flow network are modified in each iteration, the expected change at this iteration for entries not on this cycle is 0, so the expected change in corresponding entries in $P$ is 0. For those fractional edges that are modified, the probabilities are picked so that the expected adjustment in each iteration is 0. Fractional edges that are adjusted multiple times will have a variety of intermediate adjustment probabilities; still, because our procedure keeps the expected change at 0 in each iteration, the compound probabilities will also keep the expected change at 0.

\subsection{Proof of \Cref{thm:static}}\label{appendix: Cox}

\begin{proof}[Proof of \Cref{thm:static}]

Without monotonicity, the sequence of tables does not matter for the solution, and the problem is much simplified. That is, the problem earlier characterized using $Y^t$ is now reduced to $X^t$. We can even ignore the superscript and consider the problem of rounding a fractional assignment table $X$ to an integral one $\bar{X}$.
We will use \cite{cox1987constructive}'s controlled rounding algorithm that takes a fractional assignment table as input and generates an integral assignment table as output such that each fractional entry is rounded up or down to an adjacent integer. The probability of arriving at an integer assignment table will be such that the solution is ex ante proportional. To make the algorithm easier to understand, after each step, we demonstrate the algorithm on an example depicted in \Cref{fig:single prop 1}.\\

\begin{head}
\textbf{Cox's controlled rounding algorithm}

\noindent \textbf{Step 0:} Given a fractional assignment table $X$, we construct an extended table $V$ by adding an extra row to table $X$. The entry added to column $c$ is $\ceil{x_{\uni,c}}-x_{\uni,c}$, so it is zero when the column total is already an integer and otherwise equals one minus its fractional part. The extra row's total, $\sum_{c \in C \cup \{tot\}} \bigl(\lceil x_{u,c}\rceil - x_{u,c}\bigr)$, is an integer because $\sum_{c \in C} x_{u,c} = x_{u,tot}$ is an integer. Hence all row and column totals of $V$ are integral.

In our example, shown in \Cref{fig:single prop 1}, table $V$ is equivalent to table $X$ except for the last row. Adding this extra row makes the column totals integers.

\begin{figure}[!htb]
  \centering
  \begin{subfigure}{.35\linewidth}
    \centering
    \textbf{$X$} = %
    \begin{tabular}{ccc|c}
        0.5 & 0.5 & 1 & 2\tabularnewline
        0.25 & 0.25 & 0.5 & 1\tabularnewline
        0.75 & 0.75 & 1.5 & 3\tabularnewline
        \hline 
        1.5 & 1.5 & 3 & 6\tabularnewline
    \end{tabular}
    \caption{FRACTIONAL ASSIGNMENT TABLE}
  \end{subfigure}%
  \hspace{5em}%
  \begin{subfigure}{.35\linewidth}
    \centering
    \textbf{$V$} = %
    \begin{tabular}{ccc|c}
        0.5 & 0.5 & 1 & 2\tabularnewline
        0.25 & 0.25 & 0.5 & 1\tabularnewline
        0.75 & 0.75 & 1.5 & 3\tabularnewline
        0.5 & 0.5 & 0 & 1\tabularnewline
        \hline 
        2 & 2 & 3 & 7\tabularnewline
    \end{tabular}
    \caption{EXTENDED TABLE}
  \end{subfigure}%
  \caption{STEP 1 OF PROCEDURE}
  \label{fig:single prop 1}
\end{figure}

The rounding procedure involves iterative adjustment of the fractional internal entries of table $V$ until integers have replaced all fractions.

\noindent \textbf{Step 1:} If table $V$ contains no fractions, then skip to Step 7.

\noindent \textbf{Step 2:} Choose any fraction $v_{ij}$ in table $V$. At $(i, j)$, begin an alternating row-column (or column-row) path of fractions $C$. Because the total entries of $V$ are integers, each row or column containing fractions must contain at least two. Therefore, we may assume that the path of fractions contains at least four entries. At some stage, $C$ either becomes a cycle containing $(i, j)$ or else $C$ can be terminated the first time it returns to any row or column previously visited, in which case a cycle is formed that begins at the first intersection of this row or column with $C$ and ends at $C$'s terminal entry. Let $L$ denote the first cycle formed from $C$. 

\begin{figure}[!htb]
\centering
{\textbf{$V$}= %
\begin{tabular}{ccc|c}
\tikzmark{la11}\text{0.5}\tikzmark{ra11} & \tikzmark{la12}\text{0.5}\tikzmark{ra12} & \text{1} & 2\tabularnewline
\text{0.25} & \tikzmark{la22}\text{0.25}\tikzmark{ra22} & \tikzmark{la23}\text{0.5}\tikzmark{ra23} & 1\tabularnewline
\tikzmark{la31}\text{0.75}\tikzmark{ra31} & \tikzmark{la32}\text{0.75}\tikzmark{ra32} & \tikzmark{la33}\text{1.5}\tikzmark{ra33} & 3\tabularnewline
\tikzmark{la41}\text{0.5}\tikzmark{ra41} & \tikzmark{la42}\text{0.5}\tikzmark{ra42} & \text{0} & 1\tabularnewline
\hline 
2 & 2 & 3 & 7\tabularnewline
\end{tabular}%
  \begin{tikzpicture}
    [
      remember picture,
      overlay,
      -latex,
      color=black,
      yshift=1cm,
      shorten >=0.5pt,
      shorten <=1pt,
    ]

    \draw ({pic cs:ra11}) to [bend left] ({pic cs:la12});

    \draw ({pic cs:ra12}) to [bend left] ({pic cs:ra22});

    \draw ({pic cs:ra22}) -- ({pic cs:la23});
    \draw ({pic cs:ra23}) to [bend left] ({pic cs:ra33});
    \draw ({pic cs:la33}) -- ({pic cs:ra32});
    \draw ({pic cs:ra32}) to [bend left] ({pic cs:ra42});
    \draw ({pic cs:la42}) -- ({pic cs:ra41});
    \draw ({pic cs:la41}) to [bend left] ({pic cs:la11});

  \end{tikzpicture}
}
  \caption{STEP 2 OF PROCEDURE}
  \label{fig:single prop 1-2}
\end{figure}

In our example, shown in \Cref{fig:single prop 1-2}, a cycle of fractions is $(i_1, j_1) \rightarrow (i_1, j_2) \rightarrow (i_2, j_2) \rightarrow (i_2, j_3) \rightarrow (i_3, j_3) \rightarrow (i_3, j_2) \rightarrow (i_4, j_2) \rightarrow (i_4, j_1) \rightarrow (i_1, j_1)$.

\noindent \textbf{Step 3:} Choose any $\left(i_1, j_1\right)$ on $L$. Denote the members of $L$ by $\left(i_1, j_1\right),\left(i_1, j_2\right)$,$\left(i_2, j_2\right),$ $\left(i_2, j_3\right), \ldots, \ldots,\left(i_\kappa, j_{\kappa+1}\right)=$ $\left(i_\kappa, j_1\right)$, where $\kappa$ is the length of the cycle. Let
$$
\begin{aligned}
& d_{-}=\min _{1 \leq m \leq \kappa}\left[v_{i_m, j_m}-\floor{v_{i_m, j_m}},\ \ceil{v_{i_m, j_{m+1}}}-v_{i_m, j_{m+1}}\right] \\
& d_{+}=\min _{1 \leq m \leq \kappa}\left[\ceil{v_{i_m, j_m}}-v_{i_m, j_m},\ v_{i_m, j_{m+1}}-\floor{v_{i_m, j_{m+1}}}\right] .
\end{aligned}
$$
Note that both $d_{-}$ and $d_{+}$ lie strictly between 0 and 1.

\noindent \textbf{Step 4.} Select $d_{-}$ with probability $p_{-}=\frac{d_{+}}{d_{-}+d_{+}}$ or select  $d_{+}$ with probability $p_{+}=\frac{d_{-}}{d_{-}+d_{+}}$.

\noindent \textbf{Step 5a.} If $d_{+}$ is selected, move $d_{+}$ units into the $\left(i_1, j_1\right)$ cell around $L$ on an alternating $(+,-)$ path, namely,
$$
\begin{array}{ll}
\text { transform } v_{i_m, j_m} & \text { to } v_{i_m, j_m}+d_{+}, \\
\text {transform } v_{i_m, j_{m+1}} & \text { to } v_{i_m, j_{m+1}}-d_{+} .
\end{array}
$$

Return to Step 1.

This amounts to raising the odd edges and reducing the even edges at the same rate until at least one edge reaches an integer value. In our example, the odd edges rise by 0.5 and even edges reduce by 0.5 ($d_{+}=0.5$). The resulting table $V_{+}$ is shown in \Cref{fig:single prop 1-3}. 

\noindent \textbf{Step 5b.} If $d_{-}$ is selected, move $d_{-}$ units out of the $(i_1,j_1)$ cell around $L$ on an alternating $(-,+)$ path, namely, along $L$
$$
\begin{array}{ll}
\text { transform } v_{i_m, j_m} & \text { to } v_{i_m, j_m}-d_{-}, \\
\text {transform } v_{i_m, j_{m+1}} & \text { to } v_{i_m, j_{m+1}}+d_{-} .
\end{array}
$$

Return to Step 1.

This amounts to raising even edges and reducing odd edges at the same rate until at least one edge reaches an integer value. In our example, the even edges rise by 0.25, and odd edges reduce by 0.25 ($d_{-}=0.25$). The resulting table $V_{-}$ is shown in \Cref{fig:single prop 1-3}. 

\begin{figure}[!htb]
  \centering
  \begin{subfigure}{.4\linewidth}
    \centering
    \textbf{$V_{+}$} = %
    \begin{tabular}{ccc|c}
        1 & 0 & 1 & 2\tabularnewline
        0.25 & 0.75 & 0 & 1\tabularnewline
        0.75 & 0.25 & 2 & 3\tabularnewline
        0 & 1 & 0 & 1\tabularnewline
        \hline 
        2 & 2 & 3 & 7\tabularnewline
    \end{tabular}
  \end{subfigure}%
  \hspace{1em}%
  \begin{subfigure}{.4\linewidth}
    \centering
    \textbf{$V_{-}$} = %
    \begin{tabular}{ccc|c}
        0.25 & 0.75 & 1 & 2\tabularnewline
        0.25 & 0 & 0.75 & 1\tabularnewline
        0.75 & 1 & 1.25 & 3\tabularnewline
        0.75 & 0.25 & 0 & 1\tabularnewline
        \hline 
        2 & 2 & 3 & 7\tabularnewline
    \end{tabular}
  \end{subfigure}%
  \caption{STEP 5 OF PROCEDURE}
  \label{fig:single prop 1-3}
\end{figure}

Table $V$ is thus decomposed into table $V_{+}$ and table $V_{-}$ where $V=\frac{1}{3} V_{+} + \frac{2}{3} V_{-}$. There are fewer fractional entries in both tables.

\noindent \textbf{Step 6:} Repeat Steps 1--5.

\noindent \textbf{Step 7:} Transform $V$ back to $X$ by deleting the extra row added in Step 0, and subtracting its integer entries from the corresponding column totals. Report $X$ as the algorithm's outcome.
\end{head}
\vspace{0.1in}
The algorithm must end in finitely many steps, at most the number of fractional entries in table $V$. For any fractional assignment $X$, Cox's algorithm outputs integral tables with probabilities such that the induced randomized solution is ex ante proportional, locally proportional, and aggregately proportional. The following lemmas establish these properties.

\begin{lemma}\label{lemma:prop1.1}
The outcome of Cox's rounding algorithm is locally and aggregately proportional.
\end{lemma}

\begin{proof}[Proof of Lemma 4]
In Step 5, each adjustment moves $d$ units around the cycle $L$ on an alternating $(+,-)$ path, so every row and column of $V$ touched by the cycle receives one $+d$ and one $-d$; hence all row and column sums are preserved. Moreover, $d^-$ and $d^+$ are chosen so that after each adjustment every entry $v_{ij}$ still satisfies $\lfloor v_{ij}\rfloor \leq v_{ij} \leq \lceil v_{ij}\rceil$. Since the algorithm terminates only when $V$ is integral (Step 1), the final table $\bar V$ rounds every entry of $V$ to an adjacent integer and has the same row and column sums as $V$.

It remains to check Step 7. For each column $c$, the extra row's initial entry is $e_c = \lceil x_{u,c}\rceil - x_{u,c} \in [0,1)$, so its final entry satisfies $\bar e_c \in \{\lfloor e_c\rfloor, \lceil e_c\rceil\} = \{0,1\}$; and if $x_{u,c}$ is integral, then $e_c = 0$ forces $\bar e_c = 0$. The reported table $\bar X$ has university-row entry $\bar x_{u,c} = \lceil x_{u,c}\rceil - \bar e_c$, which equals $x_{u,c}$ when $x_{u,c}$ is integral and otherwise lies in $\{\lfloor x_{u,c}\rfloor, \lceil x_{u,c}\rceil\}$. Hence $\bar X$ is aggregately proportional. The internal entries of $\bar X$ are those of $\bar V$, which round the internal entries of $X$ to adjacent integers; hence $\bar X$ is locally proportional. Finally, $\bar X$ is additive: $\bar x_{u,c} = (\text{column total of } \bar V) - \bar e_c   = \sum_{d\in D} \bar v_{d,c}$, and department row totals are unchanged at $\bar x_{d,\mathrm{tot}} = x_{d,\mathrm{tot}}$.
\end{proof}

\begin{lemma}\label{lemma:prop1.2}
Cox's rounding algorithm ensures that the expected adjustment $\mathbb{E}(d)$ to each entry of $V$ at each iteration is 0. 
\end{lemma}
\begin{proof}
Notice that in each iteration of Step 1-5, the probability of raising $v_{ij}$ by $d_{+}$ is $\frac{d_{-}}{d_{-}+d_{+}}$, while the probability of reducing $v_{ij}$ by $d_{-}$ is $\frac{d_{+}}{d_{-}+d_{+}}$. Therefore, the expected adjustment is $d_{+} \frac{d_{-}}{d_{-}+d_{+}} - d_{-} \frac{d_{+}}{d_{-}+d_{+}} =0$.
\end{proof}

\Cref{lemma:prop1.2} proves that in each iteration, entries of the fractional assignment table $X$ are adjusted so that (ex ante) positive and negative deviations from fractional assignments balance to yield zero deviation in expectation. Therefore, \Cref{lemma:prop1.1} and \Cref{lemma:prop1.2} prove \Cref{thm:static}.
\end{proof}

\section{Worked examples} \label{appendix:workedexamples}

\subsection{The Government's and the Court's solutions in \Cref{example:ex1}}\label{appendix:indiaexamples}

\textbf{Government's solution.} In period-1, department $d_1$ has two vacancies, determined by the 1st and 2nd points in the university roster (that is, $\sigma(1)=c_2,\; \sigma(2)=c_2$). Department $d_2$ has one vacancy, determined by the 3rd point ($\sigma(3)=c_1$). Department $d_3$ has two vacancies, determined by the 4th and 5th points ($\sigma(4)=c_2,\; \sigma(5)=c_2$). Department $d_4$ has one vacancy, determined by the 6th point ($\sigma(6)=c_1$). The period-1 reservation table is $R_{G}(Y^1)$ in \Cref{fig:court_gov}.

In period-2, department $d_1$'s two vacancies are determined by the 7th and 8th points ($\sigma(7)=c_2,\; \sigma(8)=c_2$); department $d_2$'s vacancy by the 9th point ($\sigma(9)=c_1$); department $d_3$'s two vacancies by the 10th and 11th points ($\sigma(10)=c_2,\; \sigma(11)=c_2$); and department $d_4$'s vacancy by the 12th point ($\sigma(12)=c_1$). The period-2 reservation table is $R_{G}(Y^2)$ in \Cref{fig:court_gov}. Applying the same solution for the next period yields $R_{G}(Y^3)$.

\textbf{Court's solution.} In period-1, each department starts its own roster. Department $d_1$'s two vacancies are determined by the 1st and 2nd points of its roster ($\sigma(1)=c_2,\; \sigma(2)=c_2$); department $d_2$'s vacancy by the 1st point ($\sigma(1)=c_2$); department $d_3$'s two vacancies by the 1st and 2nd points ($\sigma(1)=c_2,\; \sigma(2)=c_2$); department $d_4$'s vacancy by the 1st point ($\sigma(1)=c_2$). The period-1 reservation table is $R_{C}(Y^1)$ in \Cref{fig:court_gov}.

In period-2, department $d_1$'s two vacancies are determined by the 3rd and 4th points of its roster ($\sigma(3)=c_1,\; \sigma(4)=c_2$); department $d_2$'s vacancy by the 2nd point ($\sigma(2)=c_2$); department $d_3$'s two vacancies by the 3rd and 4th points ($\sigma(3)=c_1,\; \sigma(4)=c_2$); department $d_4$'s vacancy by the 2nd point ($\sigma(2)=c_2$). The period-2 reservation table is $R_{C}(Y^2)$ in \Cref{fig:court_gov}. Applying the same solution for the next period yields $R_{C}(Y^3)$.

\subsection{The complete impossibility construction (\Cref{example:multi prop2})}\label{appendix:impossibilityexample}

If departments have $\mathbf{q}^3=[0,0,1]$ vacancies in period-3, department $d_3$ can reserve its vacancy for either category. These two cases are shown in \Cref{fig:multi prop 2-2}.

\begin{figure}[!htb]
    \centering
    \begin{subfigure}[b]{0.32\textwidth}
        \centering
        \textbf{$X^3$} =
        \begin{tabular}{cc|c}
            0.5 & 0.5 & 1\tabularnewline
            0 & 0 & 0\tabularnewline
            1 & 1 & 2\tabularnewline
            \hline 
            1.5 & 1.5 & 3\tabularnewline
        \end{tabular}
           \caption{\centering{PERIOD-3 FRACTIONAL ASSIGNMENT}}
    \end{subfigure}%
    \hspace{0.1em}
    \begin{subfigure}[b]{0.32\textwidth}
        \centering
        \textbf{$R_1(Y^3)$} =
        \begin{tabular}{cc|c}
            0 & 1 & 1\tabularnewline
            0 & 0 & 0\tabularnewline
            2 & 0 & 2\tabularnewline
            \hline 
            2 & 1 & 3\tabularnewline
        \end{tabular}
        \caption{\centering{PERIOD-3 INTEGRAL ASSIGNMENT}}
    \end{subfigure}%
    \hspace{0.1em}
    \begin{subfigure}[b]{0.32\textwidth}  
       \centering
        \textbf{$R_2(Y^3)$} =
        \begin{tabular}{cc|c}
            0 & 1 & 1\tabularnewline
            0 & 0 & 0\tabularnewline
            1 & 1 & 2\tabularnewline
            \hline 
            1 & 2 & 3\tabularnewline
        \end{tabular}
        \caption{\centering{PERIOD-3 INTEGRAL ASSIGNMENT}}
    \end{subfigure}
  \caption{TWO CASES FOR PERIOD-3  INTEGRAL ASSIGNMENT TABLES}
  \label{fig:multi prop 2-2}
\end{figure}

For $\ell\in\{1,2\}$, let $Y_\ell^4$ denote the fractional-assignment history ending in $X_\ell^4$ in the corresponding case below.

\textbf{Case A:} Suppose that the solution is $R_1$. If the departments have $\mathbf{q}^4=[1,0,0]$ vacancies in period-4, department $d_1$ must reserve 1 vacancy for category $c_2$; otherwise the solution would violate aggregate proportionality. The period-4 tables are $X_1^4$ and $R_1(Y_1^4)$ in \Cref{fig:multi prop 2-3}.

\textbf{Case B:} Suppose that the solution is $R_2$. If the departments have $\mathbf{q}^4=[0,1,0]$ vacancies in period-4, department $d_2$ must reserve 1 vacancy for category $c_1$; otherwise the solution would violate aggregate proportionality. The period-4 tables are $X_2^4$ and $R_2(Y_2^4)$ in \Cref{fig:multi prop 2-3}.

\begin{figure}[!htb]
    \centering
     \begin{subfigure}[b]{0.475\textwidth}
        \centering
        \textbf{$X_1^4$} =
        \begin{tabular}{cc|c}
            1 & 1 & 2\tabularnewline
            0 & 0 & 0\tabularnewline
            1 & 1 & 2\tabularnewline
            \hline 
            2 & 2 & 4\tabularnewline
        \end{tabular}
        \caption{CASE 1: PERIOD-4  FRACTIONAL ASSIGNMENT}
    \end{subfigure}%
    \hspace{0.1em}
    \begin{subfigure}[b]{0.475\textwidth}
        \centering
        \textbf{$R_1(Y_1^4)$} =
        \begin{tabular}{cc|c}
            0 & 2 & 2\tabularnewline
            0 & 0 & 0\tabularnewline
            2 & 0 & 2\tabularnewline
            \hline 
            2 & 2 & 4\tabularnewline
        \end{tabular}
        \caption{CASE 1: PERIOD-4  INTEGRAL ASSIGNMENT}
    \end{subfigure}
    \vskip\baselineskip
     \begin{subfigure}[b]{0.475\textwidth}
        \centering
        \textbf{$X_2^4$} =
        \begin{tabular}{cc|c}
            0.5 & 0.5 & 1\tabularnewline
            0.5 & 0.5 & 1\tabularnewline
            1 & 1 & 2\tabularnewline
            \hline 
            2 & 2 & 4\tabularnewline
        \end{tabular}
        \caption{CASE 2: PERIOD-4  FRACTIONAL ASSIGNMENT}
    \end{subfigure}%
    \hspace{0.1em}
     \begin{subfigure}[b]{0.475\textwidth}
        \centering
        \textbf{$R_2(Y_2^4)$} =
        \begin{tabular}{cc|c}
            0 & 1 & 1\tabularnewline
            1 & 0 & 1\tabularnewline
            1 & 1 & 2\tabularnewline
            \hline 
            2 & 2 & 4\tabularnewline
        \end{tabular}
        \caption{CASE 2: PERIOD-4  INTEGRAL ASSIGNMENT}
    \end{subfigure}
  \caption{TWO CASES FOR PERIOD-4 ASSIGNMENT TABLES}
  \label{fig:multi prop 2-3}
\end{figure}

If departments have $\mathbf{q}^5=[0,0,1]$ vacancies in period-5, department $d_3$ can reserve its vacancy for either category. These cases are shown in \Cref{fig:multi prop 2-4}.

\begin{figure}[!htb]
    \centering
    \begin{subfigure}[b]{0.32\textwidth}
        \centering
        \textbf{$X_1^5$} =
        \begin{tabular}{cc|c}
            1 & 1 & 2\tabularnewline
            0 & 0 & 0\tabularnewline
            1.5 & 1.5 & 3\tabularnewline
            \hline 
            2.5 & 2.5 & 5\tabularnewline
        \end{tabular}
        \caption{\centering{PERIOD-5  FRACTIONAL ASSIGNMENT}}
    \end{subfigure}%
    \hspace{0.1em}
    \begin{subfigure}[b]{0.32\textwidth}
        \centering
        \textbf{$R_{1.1}(Y_1^5)$} =
        \begin{tabular}{cc|c}
            0 & 2 & 2\tabularnewline
            0 & 0 & 0\tabularnewline
            3 & 0 & 3\tabularnewline
            \hline 
            3 & 2 & 5\tabularnewline
        \end{tabular}
        \caption{\centering{PERIOD-5 INTEGRAL ASSIGNMENT}}
    \end{subfigure}%
    \hspace{0.1em}
    \begin{subfigure}[b]{0.32\textwidth}  
       \centering
        \textbf{$R_{1.2}(Y_1^5)$} =
        \begin{tabular}{cc|c}
            0 & 2 & 2\tabularnewline
            0 & 0 & 0\tabularnewline
            2 & 1 & 3\tabularnewline
            \hline 
            2 & 3 & 5\tabularnewline
        \end{tabular}
        \caption{\centering{PERIOD-5 INTEGRAL ASSIGNMENT}}
    \end{subfigure}
    \vskip\baselineskip
    
    \begin{subfigure}[b]{0.32\textwidth}
        \centering
        \textbf{$X_2^5$} =
        \begin{tabular}{cc|c}
            0.5 & 0.5 & 1\tabularnewline
            0.5 & 0.5 & 1\tabularnewline
            1.5 & 1.5 & 3\tabularnewline
            \hline 
            2.5 & 2.5 & 5\tabularnewline
        \end{tabular}
        \caption{\centering{PERIOD-5   FRACTIONAL ASSIGNMENT}}
    \end{subfigure}%
    \hspace{0.1em}
    \begin{subfigure}[b]{0.32\textwidth}
        \centering
        \textbf{$R_{2.1}(Y_2^5)$} =
        \begin{tabular}{cc|c}
            0 & 1 & 1\tabularnewline
            1 & 0 & 1\tabularnewline
            2 & 1 & 3\tabularnewline
            \hline 
            3 & 2 & 5\tabularnewline
        \end{tabular}
        \caption{\centering{PERIOD-5 INTEGRAL ASSIGNMENT}}
    \end{subfigure}%
    \hspace{0.1em}
    \begin{subfigure}[b]{0.32\textwidth}  
       \centering
        \textbf{$R_{2.2}(Y_2^5)$} =
        \begin{tabular}{cc|c}
            0 & 1 & 1\tabularnewline
            1 & 0 & 1\tabularnewline
            1 & 2 & 3\tabularnewline
            \hline 
            2 & 3 & 5\tabularnewline
        \end{tabular}
        \caption{\centering{PERIOD-5 INTEGRAL ASSIGNMENT}}
    \end{subfigure}
    
  \caption{TWO CASES FOR PERIOD-5 ASSIGNMENT TABLES}
  \label{fig:multi prop 2-4}
\end{figure}

For each case until now, the period-5 reservation for category $c_1$ in department $d_1$ is 0, and the period-5 reservation for category $c_2$ in department $d_2$ is 0. The construction extends analogously for more periods via Cases A and B of \Cref{example:multi prop2}: department $d_3$ always has one vacancy in odd periods, and in the following period either $d_1$ or $d_2$ has one vacancy, forced by aggregate proportionality to the opposite category of $d_3$'s choice.

\subsection{A 100-department example for \Cref{remark:localside}}\label{appendix:100dept}

\begin{example} \label{example:100dept}
Consider a problem with 100 departments $d_1, \ldots, d_{100}$, two categories $c_1,c_2$, and reservation scheme vector $\bm{\upalpha} = [0.5 , 0.5]$. The departments have one vacancy each in period-1.

With equal category shares, local (or aggregate) proportionality requires exactly $k$ vacancies to be assigned to each category whenever a department (or the university) has $2k$ cumulative vacancies, where $k \in \mathbb{N}$.

In period-1, aggregate proportionality requires 50 departments to reserve their vacancies for $c_1$ and the other 50 to reserve theirs for $c_2$.

In period-2, suppose that only the departments that reserved a vacancy for $c_1$ in period-1 receive one additional vacancy. Local proportionality then requires each of them to reserve the additional vacancy for $c_2$. Assigning all 50 additional vacancies to $c_2$, however, produces a university-level deviation of 25 from the fractional assignment.
\end{example}

\newpage
\section{Exhibits} \label{appendix:exhibits}

This appendix reproduces the institutional documents referred to in the main text. \Cref{appendix:exhibits-rosters} shows the roster templates prescribed by the Government of India. \Cref{appendix:exhibits-university} shows how they are operated in a university: a recruitment advertisement, which is the integral assignment table the public sees, and a departmental reservation register, which is the running account behind it. \Cref{appendix:exhibits-beyond} shows the same table structure in three other Indian public-sector recruitments discussed in \Cref{section:pattern}.

\subsection{Roster templates}\label{appendix:exhibits-rosters}

\begin{figure}[h]
    \centering
    \includegraphics[width=0.6\textwidth]{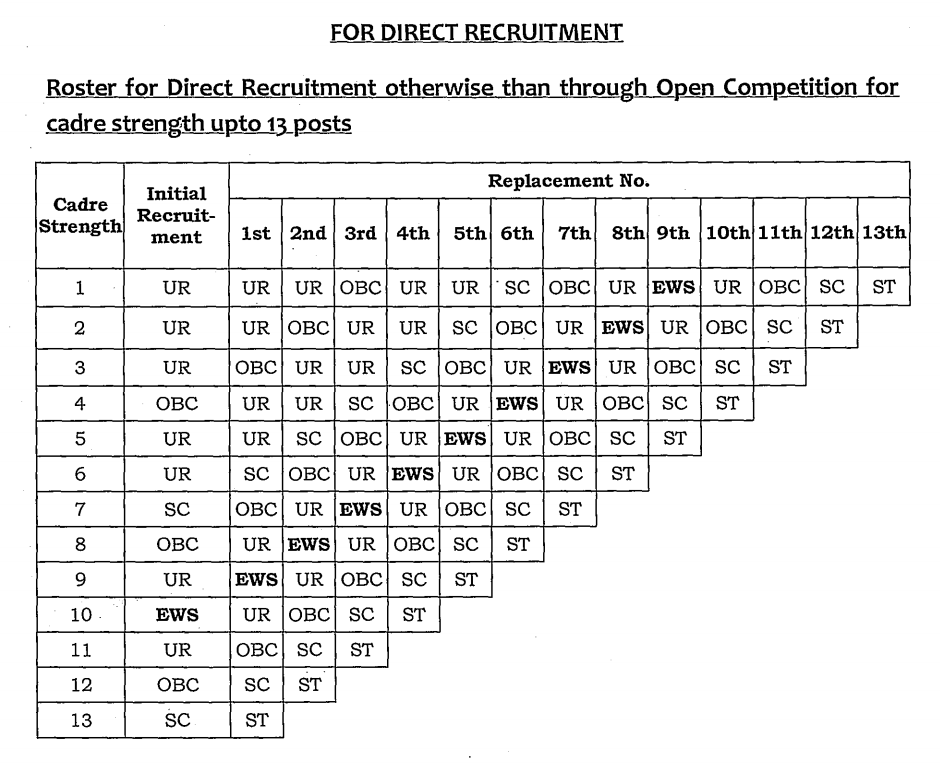}
    \floatfoot{\emph{Notes:} Template for cadres with up to thirteen posts. The first SC point is 7; the first ST point arrives only in the second cycle, at point 14. \emph{Source:} \url{https://dopt.gov.in/sites/default/files/ewsf28fT.PDF}.}
    \caption{13-POINT ROSTER PRESCRIBED BY THE GOVERNMENT OF INDIA}
    \label{fig:13pointroster}
\end{figure}

\newpage
\begin{figure}[h]
    \centering
    \includegraphics[width=0.65\textwidth]{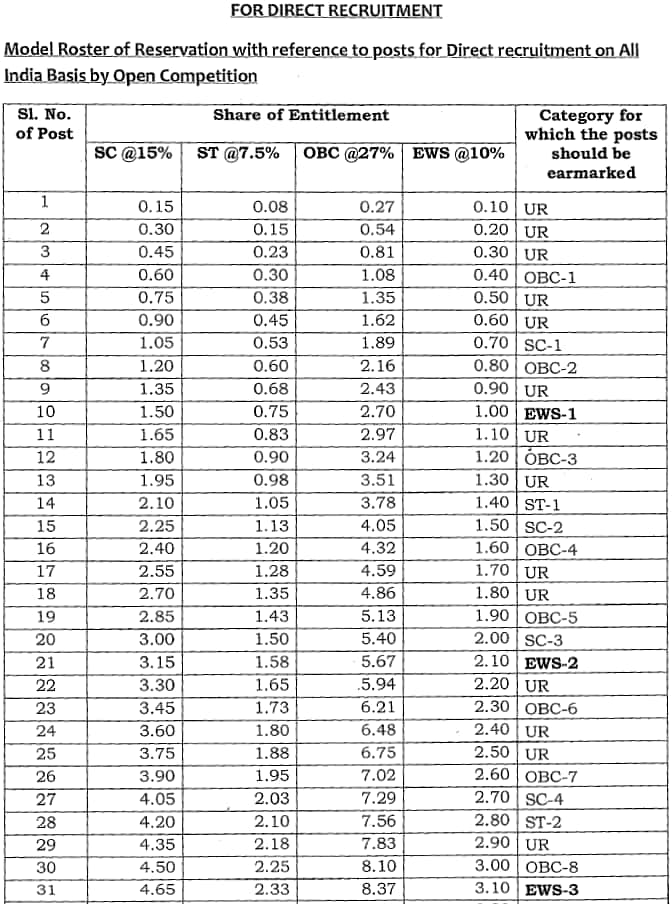}
    \caption{200-POINT ROSTER PRESCRIBED BY THE GOVERNMENT OF INDIA}
    \floatfoot{\emph{Notes:} Template for cadres with fourteen or more posts. A department with $n$ sanctioned posts uses points $1,\ldots,n$ only; each point is permanently attached to one post (\Cref{section:shortcomings}). \emph{Source:} \url{https://dopt.gov.in/sites/default/files/ewsf28fT.PDF}.}
    \label{fig:200pointroster}
\end{figure}

\newpage

\subsection{A university: advertisement and register}\label{appendix:exhibits-university}

\begin{figure}[h]
    \centering
    \includegraphics[width=0.65\textwidth]{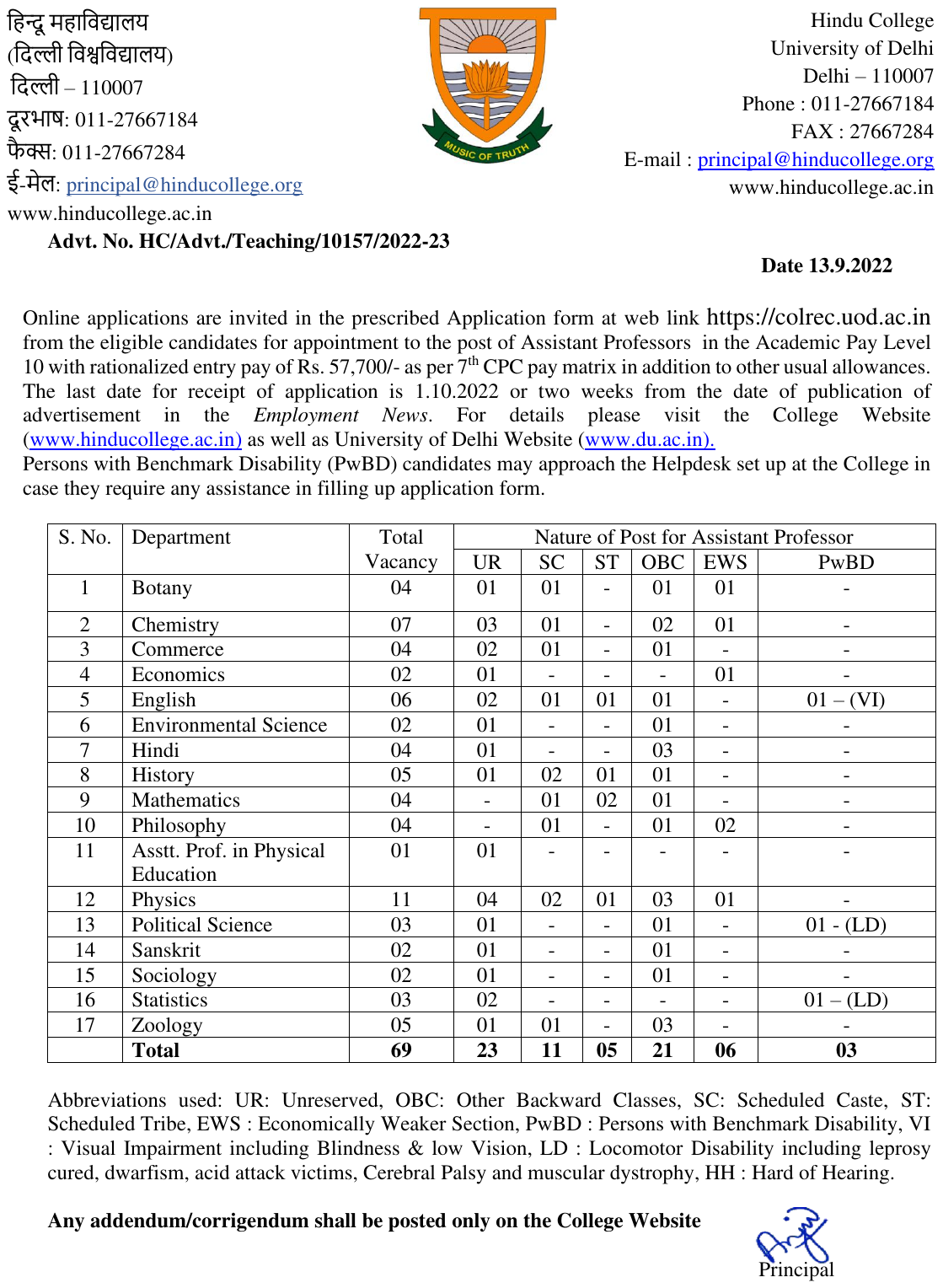}
    \floatfoot{\emph{Notes:} An integral assignment table as advertised: 69 Assistant Professor posts across 17 departments, with category counts by department and for the college as a whole. \emph{Source:} \url{https://hinducollege.ac.in/download/2022/ad/1-Advertisement\%20for\%20Assistant\%20Professors.pdf}.}
    \caption{EXAMPLE ADVERTISEMENT: HINDU COLLEGE, UNIVERSITY OF DELHI, 2022}     
    \label{fig:exampleadv}
\end{figure}

\newpage
\begin{figure}[h]
    \centering
    \includegraphics[angle=90,width=0.95\textwidth]{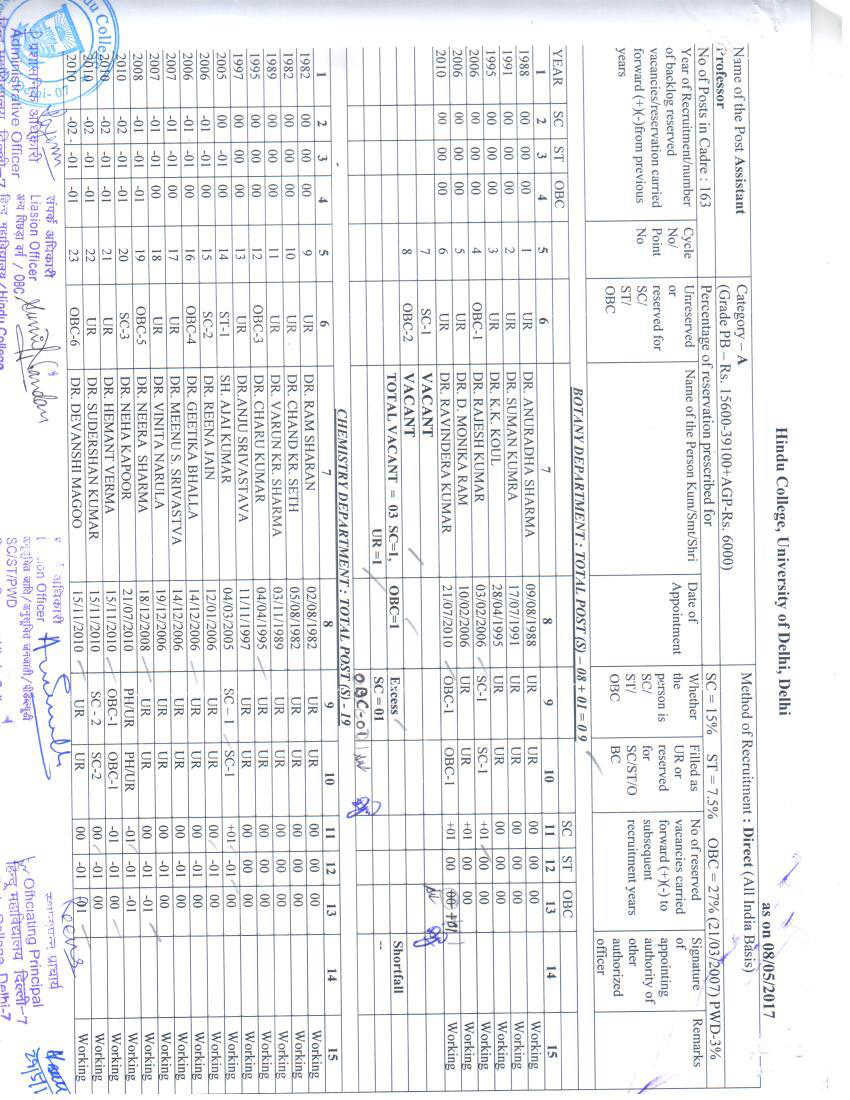} 
    \floatfoot{\emph{Notes:} Reservation register for the Assistant Professor cadre (163 posts) of Hindu College, University of Delhi, as on 8 May 2017, maintained with the department as the unit. 
    The header records the prescribed percentages (SC 15\%, ST 7.5\%, OBC 27\%, PwD 3\%). For each department the register lists, point by point: the year of recruitment, backlog reserved vacancies carried forward, the roster point and the category attached to it, the appointee and date of appointment, whether the post was filled as unreserved or reserved, and the excess or shortfall carried forward to subsequent years. Botany (9 posts) and Chemistry (19 posts) are shown. In Botany, points 1--3 are unreserved (filled 1988, 1991, 1995); point 4 is the first OBC point and point 7 the first SC point, matching \Cref{fig:200pointroster}. \emph{Source:} Publicly available roster of Hindu College, University of Delhi at \url{https://www.hinducollege.ac.in/imp-teaching-roster.aspx}}
    \caption{A RUNNING RESERVATION REGISTER (DEPARTMENT AS UNIT)}
    \label{fig:exhibit-register}
\end{figure}

\newpage

\subsection{The same table structure beyond universities}\label{appendix:exhibits-beyond}

\begin{figure}[h]
    \centering
    \includegraphics[width=0.7\textwidth]{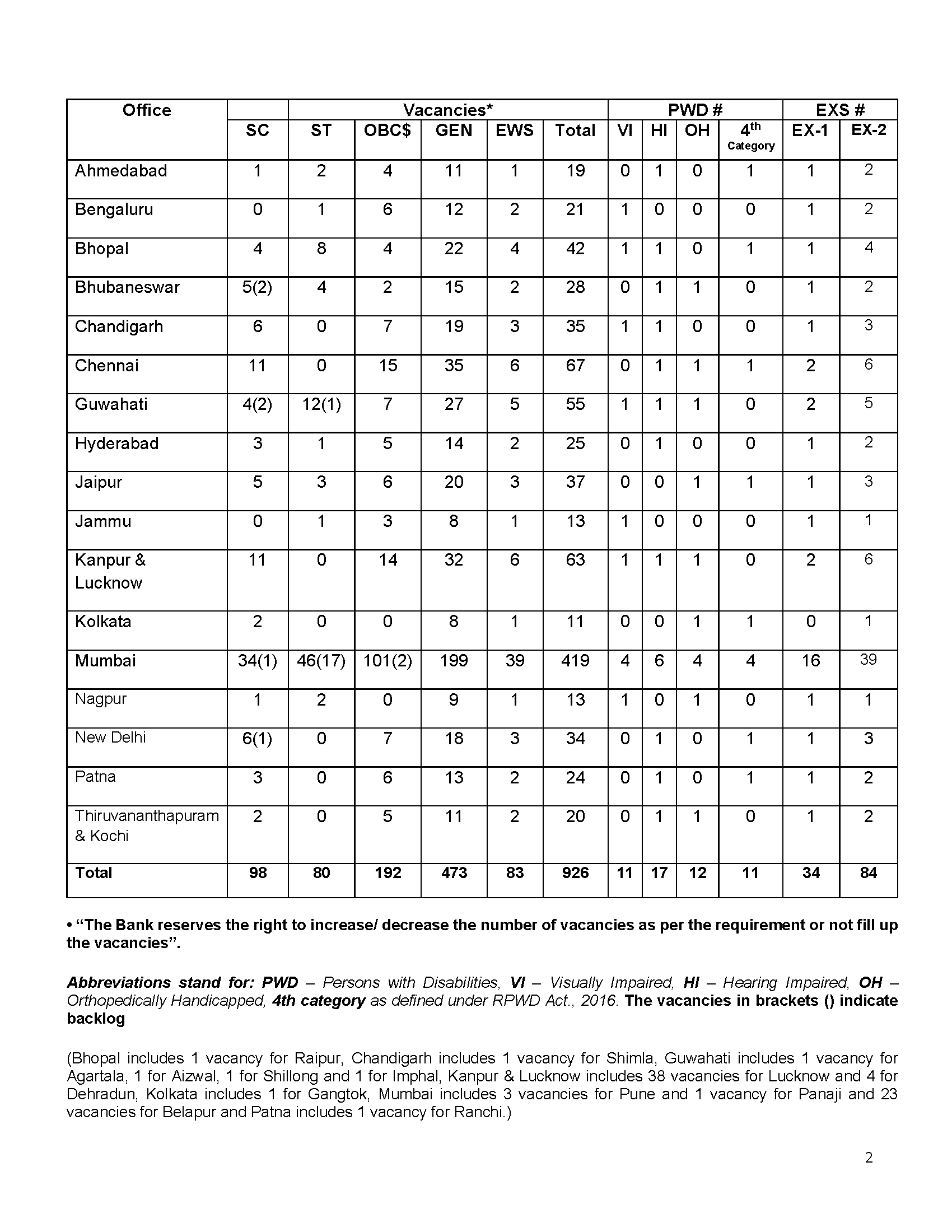} 
    \floatfoot{\emph{Notes:} Office-wise and category-wise vacancies for the post of Assistant, Reserve Bank of India: 926 posts across seventeen regional offices. Rows are offices (our departments), columns are categories, and the last row is the Bank-wide total (our university row). Bracketed entries are backlog vacancies carried forward from earlier recruitment cycles. \emph{Source:} Reserve Bank of India recruitment notification. 
    }
    \caption{RESERVE BANK OF INDIA: OFFICE-WISE DISTRIBUTION OF ASSISTANT VACANCIES}
    \label{fig:exhibit-rbi}
\end{figure}

\newpage
\begin{figure}[h]
    \centering
    \includegraphics[width=0.85\textwidth]{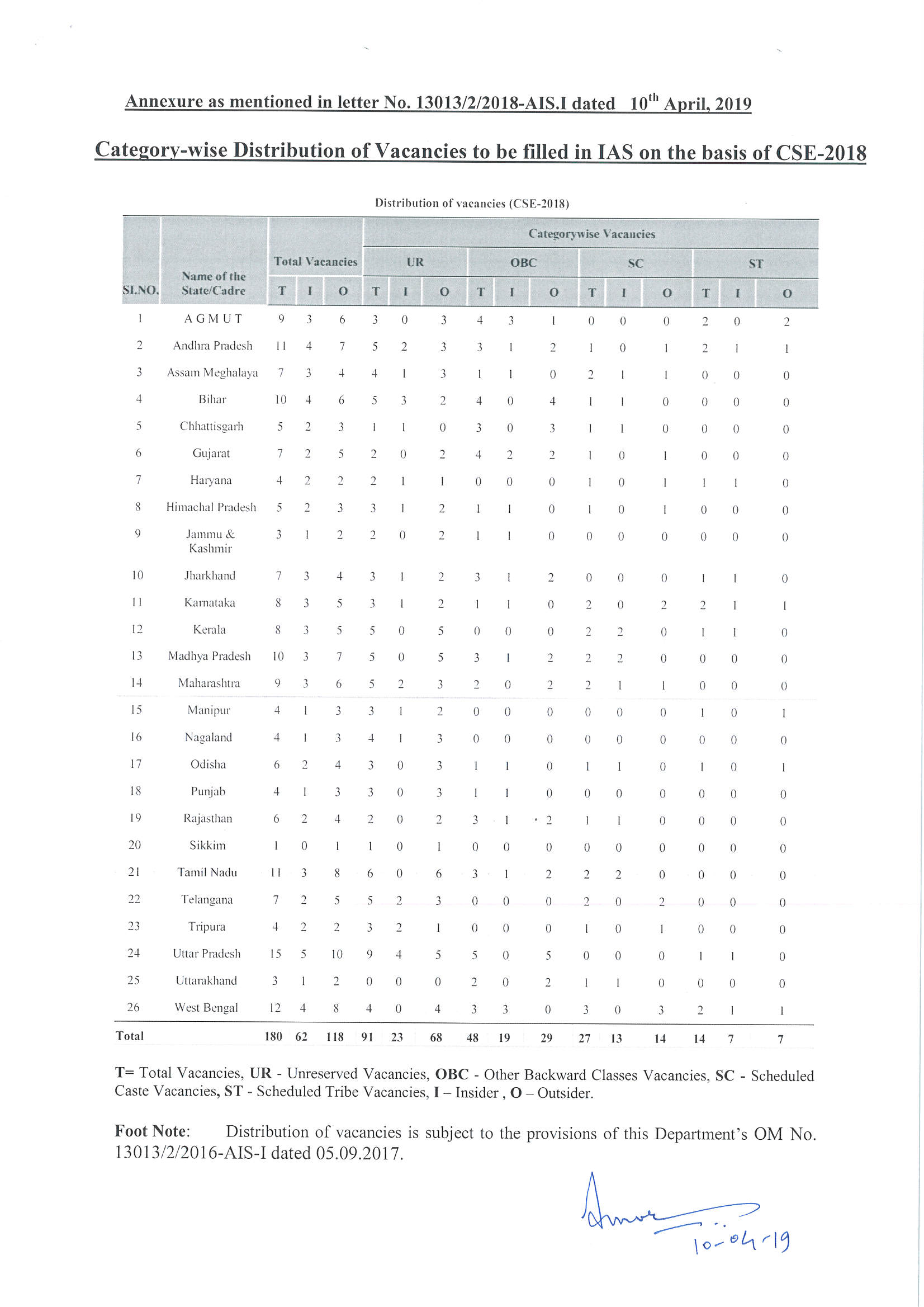} 
    \floatfoot{\emph{Notes:} Cadre-wise and category-wise distribution of 180 IAS vacancies to be filled on the basis of the Civil Services Examination 2018. Rows are the twenty-six state cadres; columns are UR, OBC, SC, and ST, each split into total (T), insider (I), and outsider (O) vacancies in a 1:2 insider--outsider ratio. Column totals track the national percentages: SC 27 (15.0\%), OBC 48 (26.7\%), ST 14 (7.8\%). \emph{Source:} Department of Personnel and Training, Annexure to letter No.\ 13013/2/2018-AIS.I dated 10 April 2019.}
    \caption{INDIAN ADMINISTRATIVE SERVICE: CADRE-WISE DISTRIBUTION OF VACANCIES, CSE-2018}
    \label{fig:exhibit-ias}
\end{figure}

\newpage
\begin{figure}[h]
    \centering
    \includegraphics[width=0.95\textwidth]{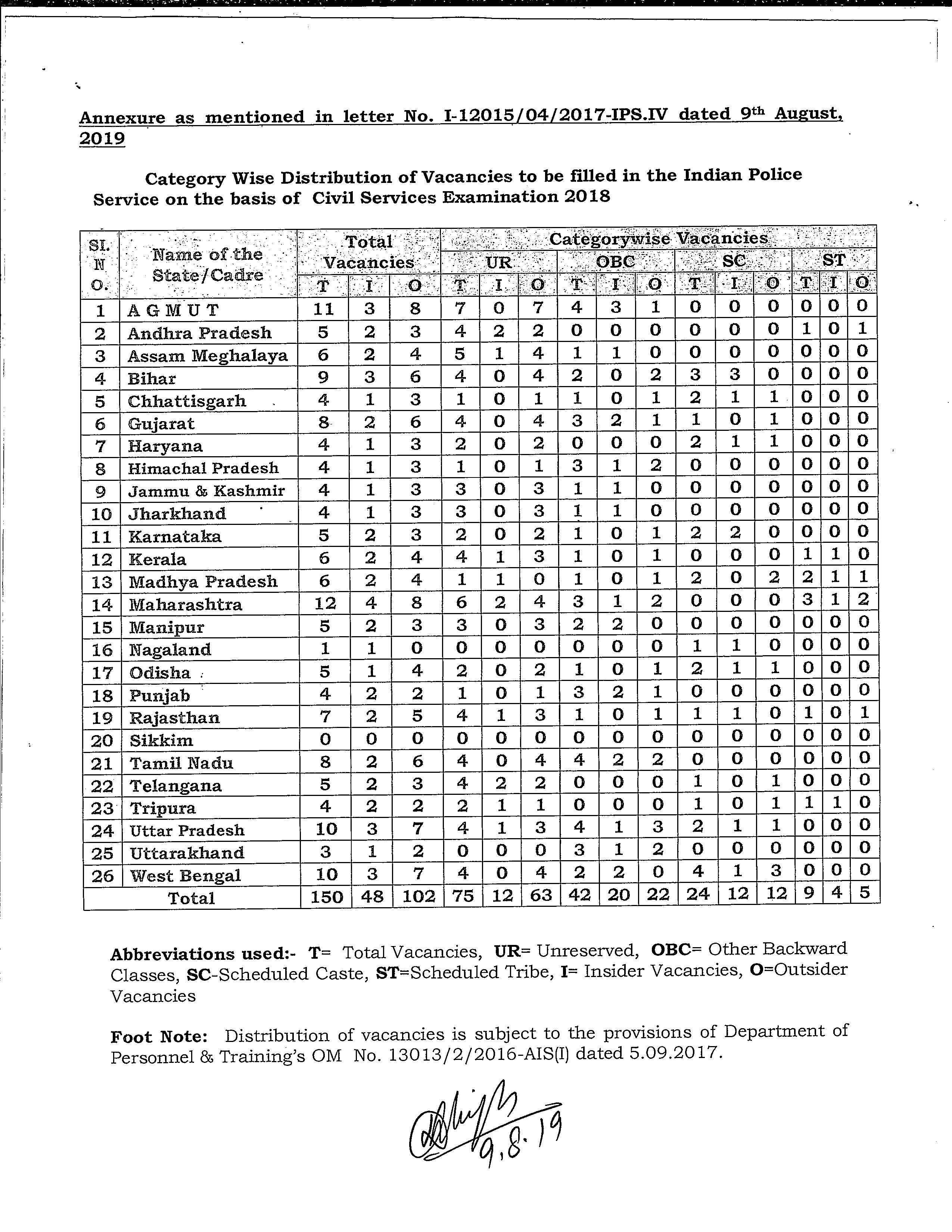} 
    \floatfoot{\emph{Notes:} Same structure for 150 IPS vacancies. Column totals: SC 24 (16.0\%), OBC 42 (28.0\%), ST 9 (6.0\%, against a fractional entitlement of 11.25). \emph{Source:} Ministry of Home Affairs, Annexure to letter No.\ I-12015/04/2017-IPS.IV dated 9 August 2019.}
    \caption{INDIAN POLICE SERVICE: CADRE-WISE DISTRIBUTION OF VACANCIES, CSE-2018}
    \label{fig:exhibit-ips}
\end{figure}

\newpage
\begin{figure}[h]
    \centering
    \includegraphics[width=0.9\textwidth]{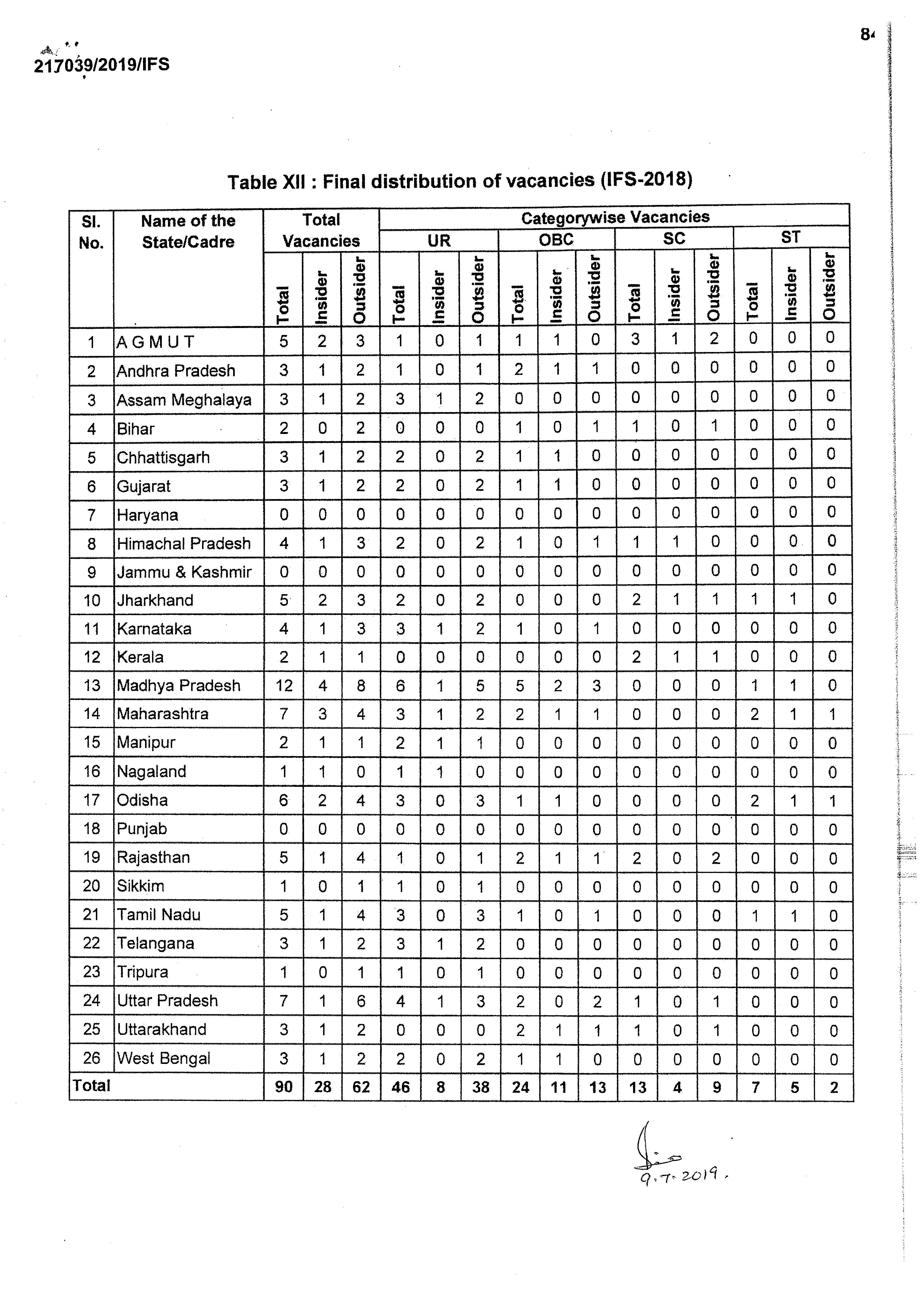} 
    \floatfoot{\emph{Notes:} Same structure for 90 Indian Forest Service vacancies. Column totals: SC 13 (14.4\%), OBC 24 (26.7\%), ST 7 (7.8\%). \emph{Source:} Ministry of Environment, Forest and Climate Change, file No.\ 217039/2019/IFS, Table XII, dated 9 July 2019.}
    \caption{INDIAN FOREST SERVICE: CADRE-WISE DISTRIBUTION OF VACANCIES, IFS-2018}
    \label{fig:exhibit-ifs}
\end{figure}


\end{document}